\documentclass[iop]{emulateapj}
\usepackage{epsfig}
\usepackage{graphics}
\shortauthors{}
\shorttitle{}

\renewcommand\deg{$^{\circ}$}
\newcommand\chandra{{\sl Chandra}}
\newcommand\xspec{\textsc{xspec}}
\newcommand\apec{\textsc{apec}}
\newcommand\projct{\textsc{projct}}
\newcommand\wabs{\textsc{wabs}}
\newcommand\vtx{v_{{\rm t}, x}}
\newcommand\vs{v_{\rm s}}
\newcommand\vrms{\sigma_{\rm t}}
\newcommand\ombv{\omega_{\rm BV}}
\newcommand\bruntvaisala{Brunt-V\"ais\"al\"a}
\newcommand\vkepler{v_{\rm K}}
\newcommand\ie{i.e.}

\makeatletter
\newcommand\@endfmtlist{\@endfmtlist}
\newcommand\@fmtrecur{\ifx\nextfmtchar\@endfmtlist 
    \ifsuperdone \else ^{\mysuper}\fi \let\nextfmtaction\relax
  \else \ifx\nextfmtchar. ^{\mysuper}\!.\superdonetrue \else \nextfmtchar\fi
    \let\nextfmtaction\@fmtdeclist \fi 
  \nextfmtaction}
\newcommand\@fmtdeclist{\afterassignment\@fmtrecur\let\nextfmtchar=}
\newcommand\dofmt[1]{\newif\ifsuperdone\superdonefalse
  \@fmtdeclist#1\@endfmtlist}
\newcommand\supersec{\rm s}
\newcommand\hhmmss[3]{\let\mysuper\supersec #1^{\rm h}\,
  #2^{\rm m}\, \dofmt{#3}}
\newcommand\superarcsec{\prime\prime}
\newcommand\ddmmss[3]{\let\mysuper\superarcsec #1^\circ\, #2'\, \dofmt{#3}}
\def\dissecthms#1 #2 #3\@nil{{#1}{#2}{#3}}
\newcommand\fmtra[1]{$\expandafter\hhmmss\dissecthms#1\@nil$}
\newcommand\fmtdec[1]{$\expandafter\ddmmss\dissecthms#1\@nil$}
\newcommand\fmtradec[2]{$(\expandafter\hhmmss\dissecthms#1\@nil,
  \expandafter\ddmmss\dissecthms#2\@nil)$}
\makeatother

\begin{document}

\title{Deep {\sl Chandra} Observations of Abell 2199:  The Interplay
  Between Merger-Induced Gas Motions and Nuclear Outbursts in a Cool
  Core Cluster}  
\author{Paul~E.~J.~Nulsen\altaffilmark{1},
  Zhiyuan~Li\altaffilmark{1,2}, William~R.~Forman\altaffilmark{1},
  Ralph~P.~Kraft\altaffilmark{1}, Dharam~V.~Lal\altaffilmark{1,3}, 
  Christine~Jones\altaffilmark{1}, Irina~Zhuravleva\altaffilmark{4,5},
  Eugene~Churazov\altaffilmark{6}, Jeremy~S.~Sanders\altaffilmark{7,8},
  Andrew~C.~Fabian\altaffilmark{8}, Ryan~E.~Johnson\altaffilmark{9},
  Stephen~S.Murray\altaffilmark{1,10}} 
\altaffiltext{1}{Harvard-Smithsonian Center for Astrophysics, 60
  Garden Street, Cambridge, MA 02138}
\altaffiltext{2}{UCLA Physics \& Astronomy, BOX 951547 3-351 PAB, Los
  Angeles, CA 90095}
\altaffiltext{3}{National Centre for Radio Astrophysics, NCRA-TIFR,
  Pune University Campus, Ganeshkhind P.O., Pune 411 007, India}
\altaffiltext{4}{Kavli Institute for Particle Astrophysics and
  Cosmology (KIPAC), Stanford University, 452 Lomita Mall, Stanford,
  CA 94305}
\altaffiltext{5}{Department of Physics, Stanford University, 452
  Lomita Mall, Stanford, CA 94305} 
\altaffiltext{6}{Max-Planck-Institut f\"ur Astrophysik,
  Karl-Schwarzschild-Str. 1, D-85748 Garching, Germany}
\altaffiltext{7}{Max-Planck-Institut f\"ur extraterrestrische Physik,
  Giessenbachstrasse 1, 85748 Garching, Germany}
\altaffiltext{8}{University of Cambridge, Institute of Astronomy,
  Madingley Road, Cambridge CB3 0HA, UK}
\altaffiltext{9}{Denison University, Department of Physics and
  Astronomy, Granville, OH 43023}
\altaffiltext{10}{Johns Hopkins University, 3400 North Charles St.,
  Baltimore, MD 21205}

\begin{abstract}

We present new \chandra{} observations of Abell~2199 that show
evidence of gas sloshing due to a minor merger, as well as impacts of
the radio source, 3C~338, hosted by the central galaxy, NGC~6166, on
the intracluster gas.  The new data are consistent with previous
evidence of a Mach $\simeq 1.46$ shock $100''$ from the cluster
center, although there is still no convincing evidence for the
expected temperature jump.  Other interpretations of this feature are
possible, but none is fully satisfactory.  Large scale asymmetries,
including enhanced X-ray emission $200''$ southwest of the cluster
center and a plume of low entropy, enriched gas reaching $50''$ to the
north of the center, are signatures of gas sloshing induced by core
passage of a merging subcluster about 400 Myr ago.  An association
between the unusual radio ridge and low entropy gas are consistent
with this feature being the remnant of a former radio jet that was
swept away from the AGN by gas sloshing.  A large discrepancy between
the energy required to produce the $100''$ shock and the enthalpy of
the outer radio lobes of 3C~338 suggests that the lobes were formed by
a more recent, less powerful radio outburst.  Lack of evidence for
shocks in the central $10''$ indicates that the power of the jet now
is some two orders of magnitude smaller than when the $100''$ shock
was formed.

\end{abstract}
\keywords{X-rays: galaxies: clusters -- galaxies: clusters: individual
  (Abell~2199) -- intergalactic medium -- cooling flows}

\section{Introduction} {\label{sec:intro}}

The nearby rich cluster Abell 2199 (hereafter A2199) is a prototypical
``cooling flow'' cluster.  On large scales, the morphology of its
intracluster medium (ICM) is smooth and highly symmetric, as revealed
by {\sl Einstein} \citep{fj82}, {\sl ROSAT} \citep{ssj98}, and {\sl
  ASCA} \citep{mvf99} observations.  Embedded in the cool cluster core
is the cD galaxy NGC~6166, which hosts the spectacular and unusual
radio source 3C\,338 \citep{bsw83}.  Using {\sl ROSAT}/HRI and {\sl
  VLA} observations, \citet{oe98} examined the disturbed morphology of
the core gas and concluded that nuclear outbursts have likely
deposited a significant amount of energy in the core, thus
dramatically modifying the gas dynamics.  Through analysis of early
\chandra{} observations, \citet{jaf02} confirmed and extended our
understanding of the gas dynamics in A2199, revealing an X-ray cavity
associated with the eastern radio lobe and quantifying the gas heating
required to maintain the cool core in a steady state.  From these
\chandra{} observations, \citet{sf06} found a surface brightness
discontinuity in the gas $100''$ from the cluster center, which they
attributed to an isothermal shock driven into the gas by the
supersonic inflation of the 3C\,338 radio lobes.

In this paper we present results from new \chandra{} observations of
A2199, providing approximately four times the exposure and a
larger field 
of view than the earlier observations.  Section~\ref{sec:data}
describes the data preparation. Section~\ref{sec:overview} presents an
overview of the X-ray and radio structures.
Section~\ref{sec:discussion} discusses the effects of sloshing and
outbursts from the active nucleus as the origin of the various
structures summarized in section~\ref{sec:overview}.
Section~\ref{sec:summary} summarizes the results.  We adopt a
luminosity distance of 127 Mpc ($1' = 34.8$ kpc) for A2199, based on
the redshift of NGC 6166 \citep[$z = 0.0304$;][]{ddc91}.
Uncertainties are quoted at the 90\% confidence level, unless
otherwise stated.  All spectral fits include foreground absorption
($N_H = 9 \times 10^{19}\rm\ cm^{-2}$).

\section{Chandra and VLA Observations} {\label{sec:data}}

\subsection{Chandra Imaging Analysis Methods} 

\chandra{} observed A2199 on six occasions, with two 20 ksec ACIS-S
observations in 1999 -- 2000 and four ACIS-I observations, for a
total exposure of 120 ksec, in 2009.  A log of the \chandra{}
observations is given in Table~\ref{tab:log}.  In this work, we use
data collected by the S3 CCD in the two ACIS-S observations and data
from the I0 -- 3 CCDs in the ACIS-I observations.

All six observations were reprocessed with CIAO v4.2, using the
corresponding calibration files and following standard reduction
procedures.  The light curves indicate that the instrumental
background for each observation was fairly quiescent.  Filtering out
periods of high background using $3\sigma$ clipping with the CIAO
LC\_CLEAN
script\footnote{http://cxc.harvard.edu/ciao/ahelp/lc\_clean.html}
resulted in $\sim$10\% exposure loss (Table~\ref{tab:log}). Our
analysis was restricted to the 0.5 -- 8 keV range.  Exposure maps for
the observations were generated using energy weights determined from a
thermal plasma model (APEC), with a temperature of 4 keV and
abundances of 0.3 solar on the scale of \citet{ag89}, plus fixed
foreground absorption (WABS), appropriate for the large-scale X-ray
emission from A2199.  We also accounted for the energy-dependence in
the effective area of the S3 CCD and I CCDs, so that the count rates
given in this paper are appropriate to ACIS-I.  ``Blank sky''
data\footnote{http://cxc.harvard.edu/ciao/threads/acisbackground/}
were used to generate the sky+particle background for each
observation, after calibrating the particle background level in the 10
-- 12 keV range.  Background levels are generally less than a few
percent for the data analyzed here.  Artifacts due to photons recorded
during CCD readout were accounted for, following the technique of
\citet{mpn00}.  This procedure simulates photon events recorded during
readout for individual observations.  These ``out-of-time'' events
were combined with the ``blank sky'' events to form a fiducial
background to be subtracted from the observations.

Like almost every X-ray mission to date, \chandra's broad band
response is insensitive to the gas temperature for values in the range
of interest here.  For example, for a fixed gas emission measure,
reducing the nominal gas temperature from 4 to 3 keV would increase
the 0.5 -- 2 keV count rate in ACIS-I by $\simeq 1.3\%$, while
increasing it to 5 keV would reduce the count rate by $\simeq1.1\%$.
The effect of such temperature variations alone would be imperceptible
in our images.  With other parameters fixed, the count rate is
proportional to the emission measure, \ie, to the square of the gas
density within a fixed volume.  Away from shocks, the gas remains
close to local pressure equilibrium, so that an abrupt temperature
change is usually acompanied by an abrupt change in the density.  In
that case, the density change accompanying a temperature reduction
from 4 to 3 keV would boost the count rate by $\simeq 78\%$, while
that accompanying a temperature boost from 4 to 5 keV would decrease
the count rate by $\simeq 36\%$.  Thus, under local pressure
equilibrium, changes in the 0.5 -- 2 keV surface brightness are
dominated by the impact of the change in density.  In a weak shock,
both the temperature and the density increase, with the fractional
temperature change being the smaller of the two (to first order, by a
factor of $2/3$ in monatomic gas), so that the density change is also
the dominant cause of changes in brightness associated with shocks.
The 0.5 -- 2 keV response of \chandra{} is also quite insensitive to
the abundances.  For example, doubling the abundance from 0.3 to 0.6
solar in the thermal model used to make the exposure map would
increase the image brightness by $\simeq 10\%$.  In summary, structure
in the X-ray images is almost completely determined by structure in
the gas density.

\begin{figure*}
\centerline{
\epsfig{figure=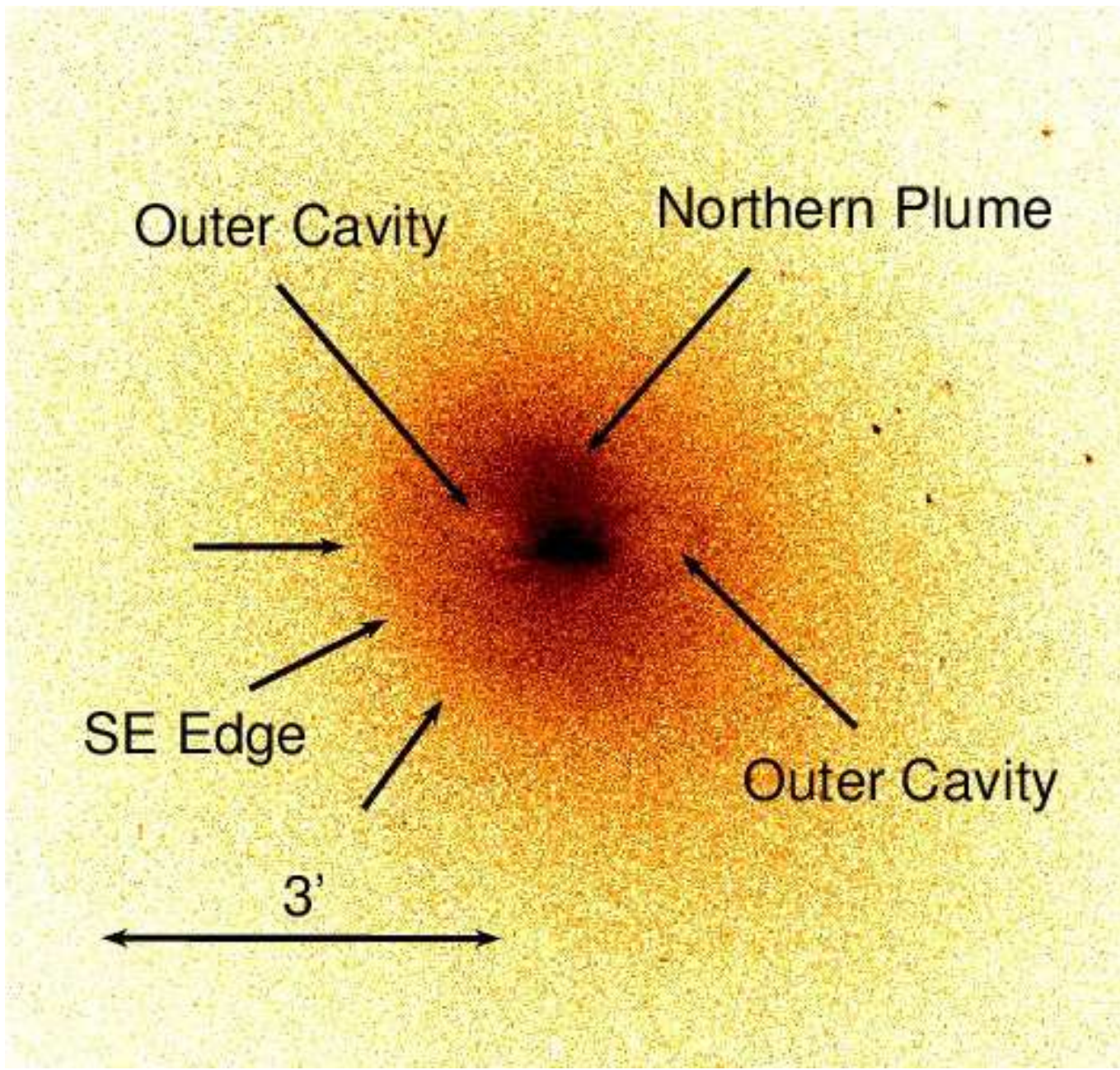,width=\textwidth,clip=}
}
\caption{Background-subtracted, exposure-corrected 0.5 -- 2 keV
  \chandra{} image of A2199.  The data are in block 2 ($\simeq 1''$
  pixels), with no further smoothing.  The image shows the most
  prominent features including the southeastern edge, the two outer
  X-ray cavities to the east and west, a plume of emission to the
  north. The image hints at asymmetries on larger scales, which can be
  seen more clearly in the resdual image of Fig.~\ref{fig:excess}.
  The surface brightness at the centers of the two outer cavities is
  approximately 20\% lower than in the regions surrounding the
  cavities, as expected if they lie close to the plane of the sky and
  are devoid of X-ray emitting gas \citep[e.g.,][]{wmn07}.  The color
  scale is logarithmic in the surface brightness.}
\label{fig:fov}
\end{figure*}

For the imaging analysis, we projected the individual count, exposure
and background maps to a common tangent point at the optical center of
NGC~6166 to produce summed images of the combined field of view. The
central $\sim 200''$ of A2199 is covered by all six observations,
whereas the central $200''$ -- $400''$ region is covered by the four
ACIS-I observations.  Point sources detected across the combined field
of view, apart from the X-ray nucleus of NGC~6166, were excluded from
further quantitative analysis.  The merged, background subtracted and
exposure corrected image is shown in Fig.~\ref{fig:fov} and discussed
below.

\begin{figure*}
\centerline{
\epsfig{figure=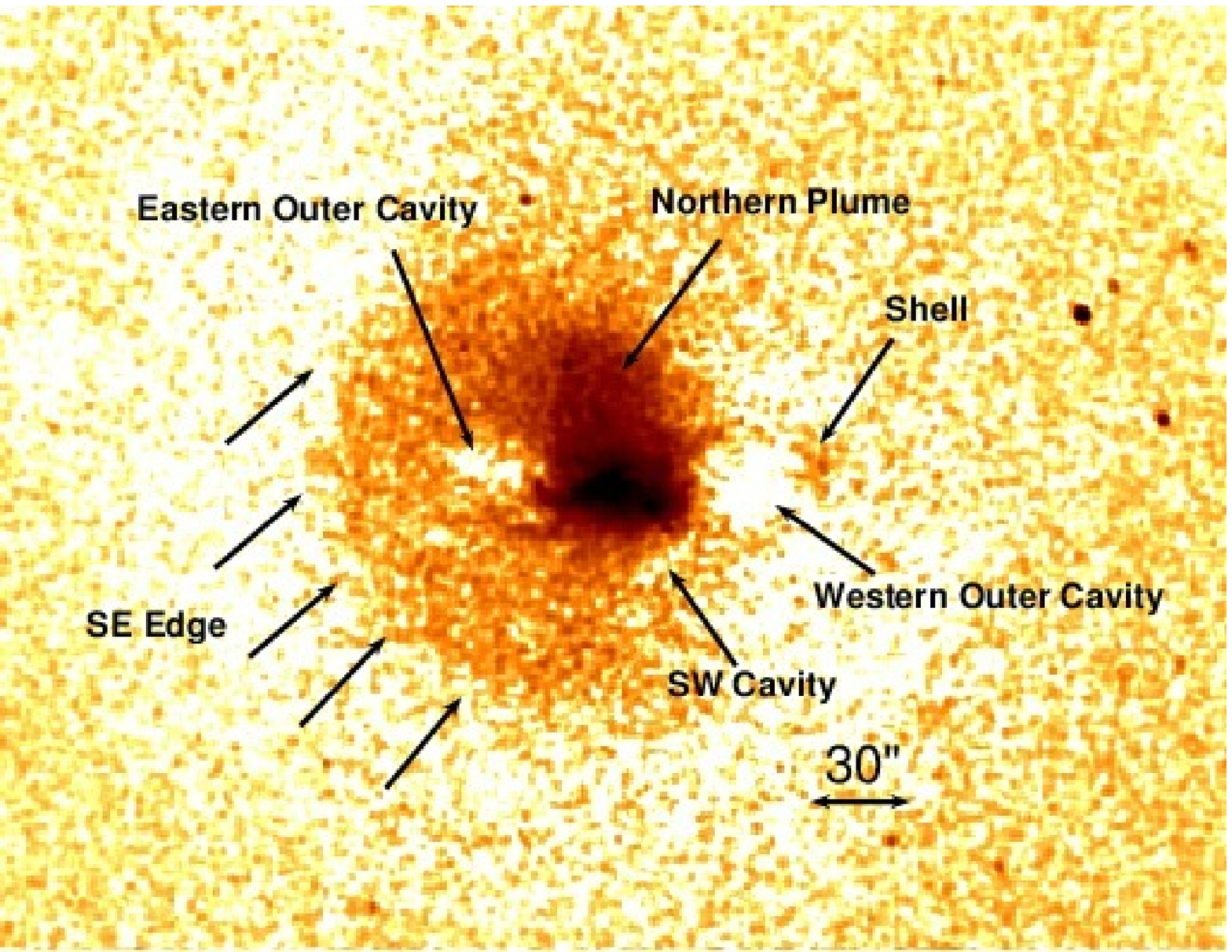,width=0.48\textwidth,clip=}
\epsfig{figure=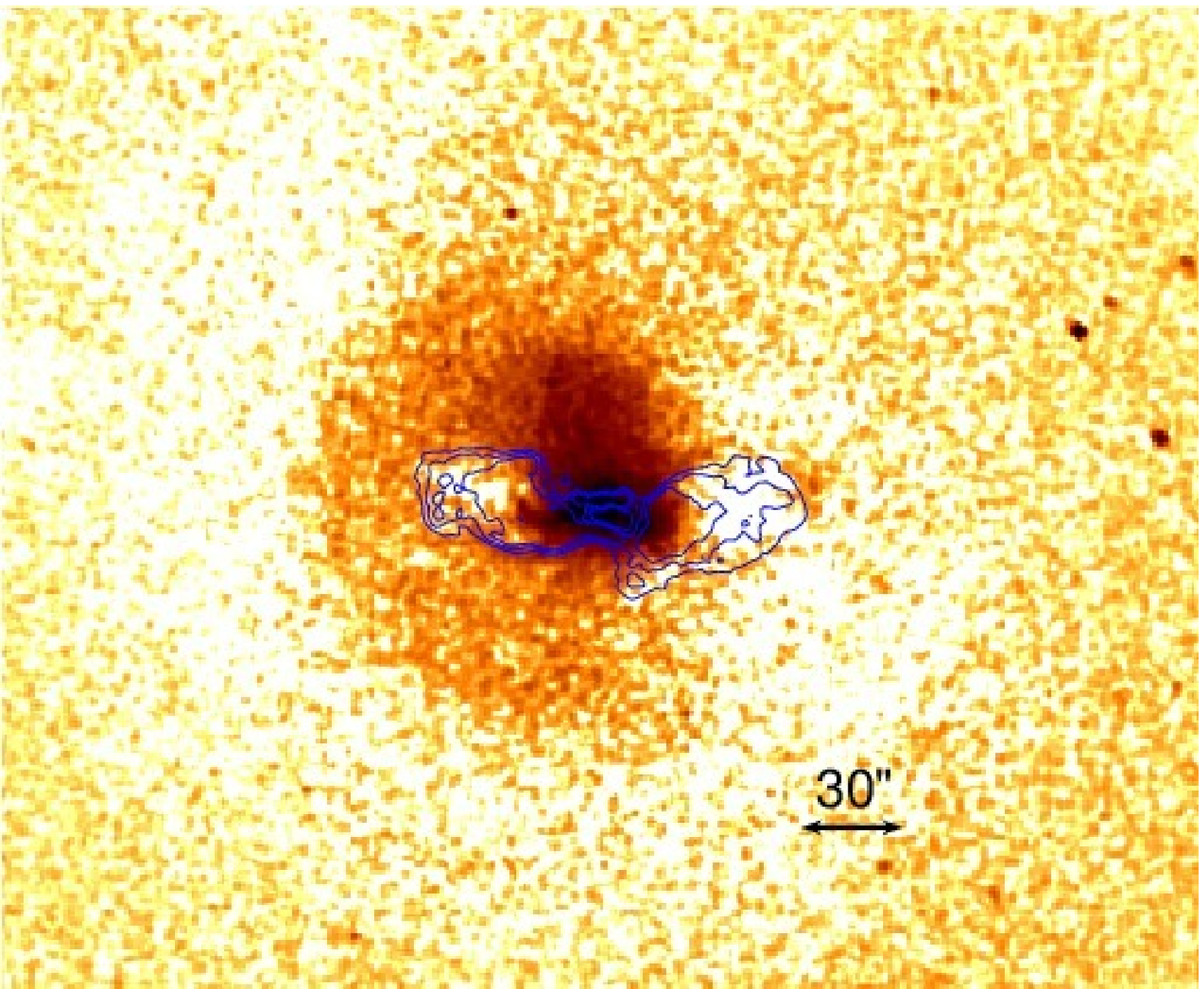,width=0.45\textwidth,clip=}
}
\caption{{\sl Left:} The 0.5 -- 2 keV residual image of the central
  $200'' \times 200''$ region, obtained by subtracting the best-fit
  elliptical $\beta$-model from the X-ray image.  The image has been
  smoothed with a Gaussian kernel of FWHM = $2''$.  Prominent features
  in the image include the SE edge, the cavities, the northern plume,
  and the shell surrounding the Western Outer cavity. {\sl Right:} The
  same image with 4.9 GHz contours overlaid (25, 50, $100\ \mu\rm
  Jy\ beam^{-1}$).  The radio emitting plasma fills the three cavities
  seen in the X-ray emission, largely excluding X-ray emitting gas
  from them. }
\label{fig:central}
\end{figure*}

\begin{figure}
\centerline{\epsfig{figure=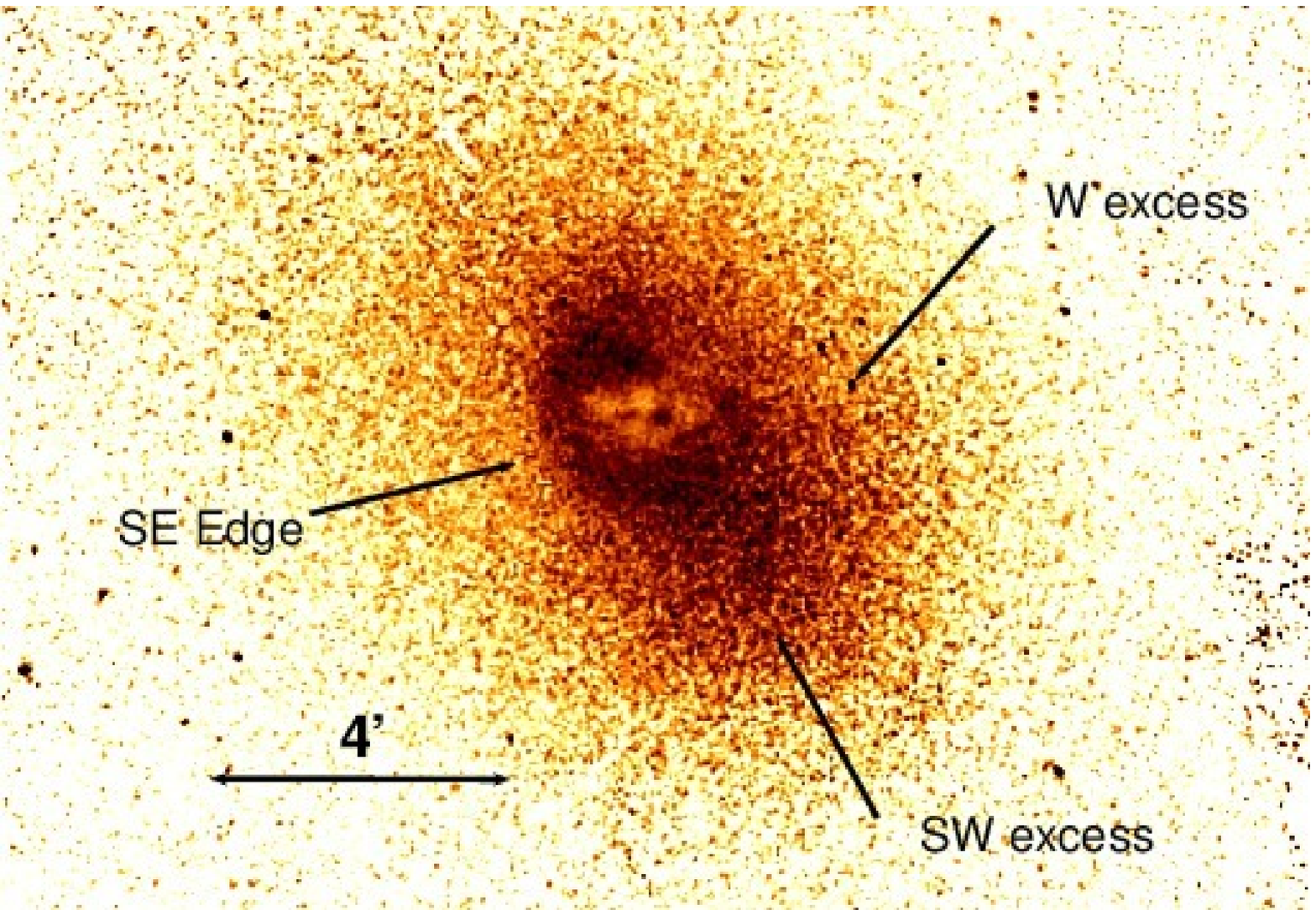,width=0.45\textwidth,clip=}}
\caption{Background subtracted, exposure corrected, 0.5 -- 2 keV
  residual \chandra{} image of A2199 after dividing by an
  azimuthally symmetric X-ray profile.  Strong excess
  emission is seen to the southwest and west.  The Southeast edge is
  very prominently seen extending over nearly 180\deg{} in azimuth.  }
\label{fig:excess}
\end{figure}

For the purpose of highlighting residual structure in the X-ray image,
the two-dimensional X-ray surface brightness distribution was fitted
with an elliptical $\beta$-model using CIAO {\sl Sherpa} and the
C-statistic to measure goodness of fit.  The best fitting parameters
depend to some extent on the region fitted (Table~\ref{tab:beta}), but
they are in reasonable agreement with previous results from {\sl
  ROSAT} \citep{oe98, ssj98}.  Parameters for the range $60''$ --
$400''$ were used to make the residual images of
Fig.~\ref{fig:central}.  To obtain the image in Fig.~\ref{fig:excess},
the X-ray image was divided by a beta model fitted to the azimuthally
averaged surface brightness profile, highlighting departures from
azimuthal symmetry.  We note that the features seen in the images are
insensitive to the underlying cluster profiles.

\begin{figure}
\centerline{
\epsfig{figure=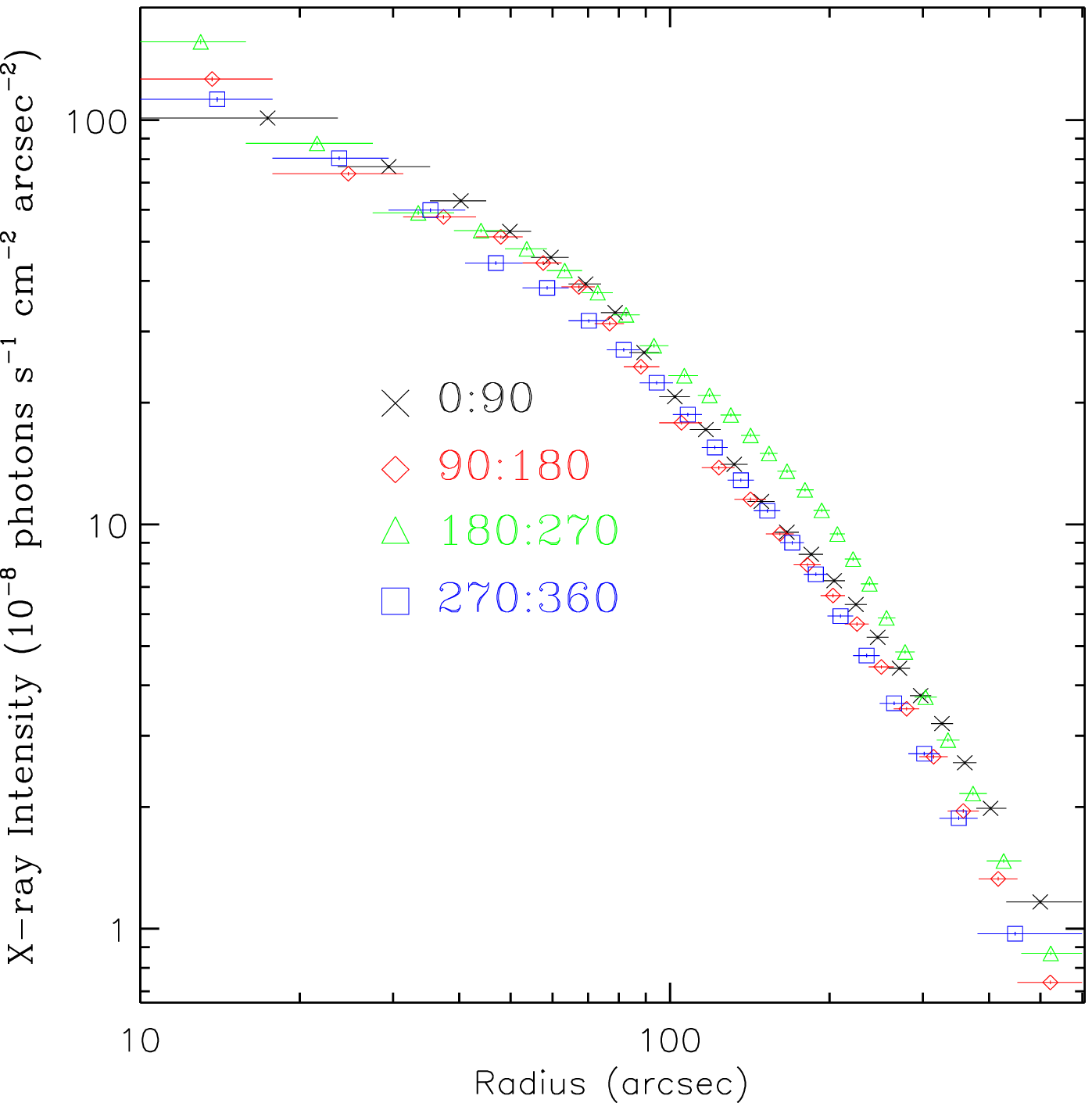,width=0.4\textwidth,angle=0,clip=}
}
\caption{Azimuthally averaged profiles in four sectors (0\deg{} --
  90\deg, 90\deg{} -- 180\deg, 180\deg{} -- 270\deg, 270\deg{} --
  360\deg{}).  The excess beyond $100''$ in the profile for the
  southwestern sector corresponds to the feature marked in
  Fig.~\ref{fig:excess}.  }
\label{fig:profiles}
\end{figure}

\subsection{Chandra Spectral Analysis Methods}

For spectral analysis of the ICM, we extracted spectra and produced
instrument response files for individual observations,
using the standard event weighting for extended regions.
Point sources were identified using the method of \citet{w04}, with a
detection limit of $2.5\times10^{-4}\rm\ ct\ s^{-1}$ (0.5 -- 8 keV),
giving 60 point source regions that were excluded from spectra.  In
most cases, spectral results were obtained by fitting spectra jointly.
However, for the fits given in Table~\ref{tab:spec}, since the
available photon counts are small, the spectra were co-added for
fitting.  The temperature and metallicity maps were also made by
fitting summed spectra.  Tests show that the differences in
the best fitting spectral parameters obtained by fitting the
co-added spectrum and fitting the individual spectra jointly are
smaller than a few percent.

\subsection{VLA Observations}

Two archival VLA datasets for 3C 338 taken at 1.5 GHz in A-array
configuration and 4.9 GHz in B-array configuration
(Table~\ref{tab:radio}) were used to make radio images for comparison
with the X-ray and to measure radio properties with high angular
resolution (Table~\ref{tab:radio2}).  The radio observations were
analyzed using AIPS (Astronomical Image Processing System).  Absolute
flux density calibration was tied to observations of 3C\,286
\citep{bg09}.

\section{X-ray, Radio, and Multiwavelength Structures}
 {\label{sec:overview}}

The multiwavelength observations, especially the \chandra{} X-ray and
VLA radio images, provide insights into the origin of the variety of
structures seen in A2199. Below we describe the structures and in
section~\ref{sec:discussion}, we discuss possible physical origins
for these structures.

\subsection{Large Scale X-ray Structures - Southwest Excess and Eastern Edge} 

The X-ray images show the presence of several large scale structures.
The primary features seen in the large scale images of
Figs.~\ref{fig:fov} -- \ref{fig:excess} include:
\begin{itemize}
\item a large scale excess to the southwest and west, most clearly
  apparent in the residual image of Fig.~\ref{fig:excess}.
\item a prominent southeastern edge (hereafter SE edge)  extending
  over azimuths\footnote{Azimuthal angles are measured eastward
    from north.} 45$^\circ$ to 180$^\circ$. This 
  feature was previously discussed by \citet{jaf02} and \citet{sf06}
\end{itemize}

To further investigate the large scale asymmetry in the gas
distribution, radial surface brightness profiles were made in four
sectors (Fig.~\ref{fig:profiles}). The asymmetry is most apparent in
the southwest quadrant, which shows an excess over the other quadrants
from $100''$ to $250''$, peaking at $\simeq 30\%$ at a radius of
$200''$.  This is the excess seen in the X-ray images to the southwest
in Fig.~\ref{fig:excess}.

\begin{figure*} [thb]
\centerline{
\epsfig{figure=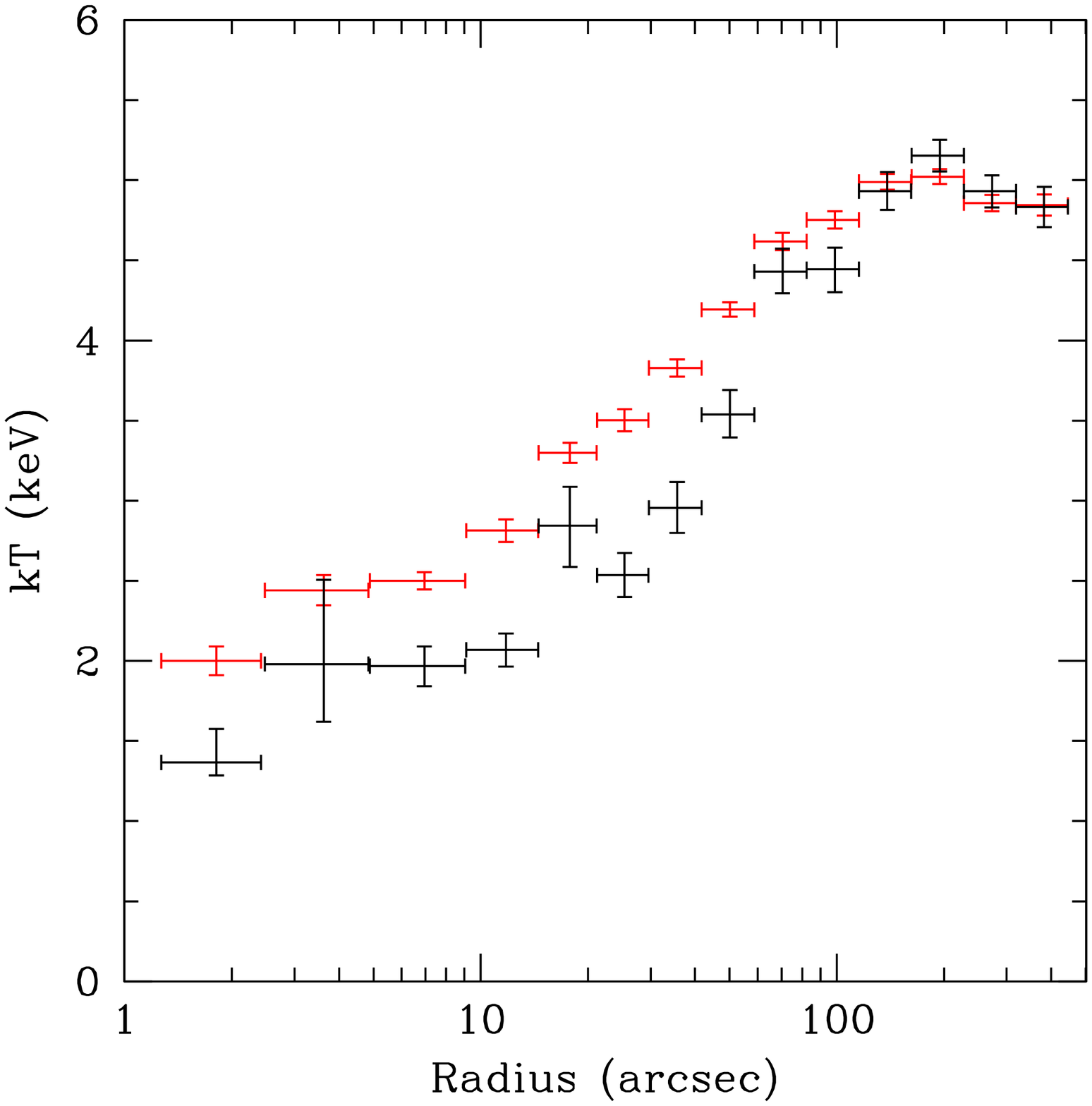,height=0.48\textwidth}
\epsfig{figure=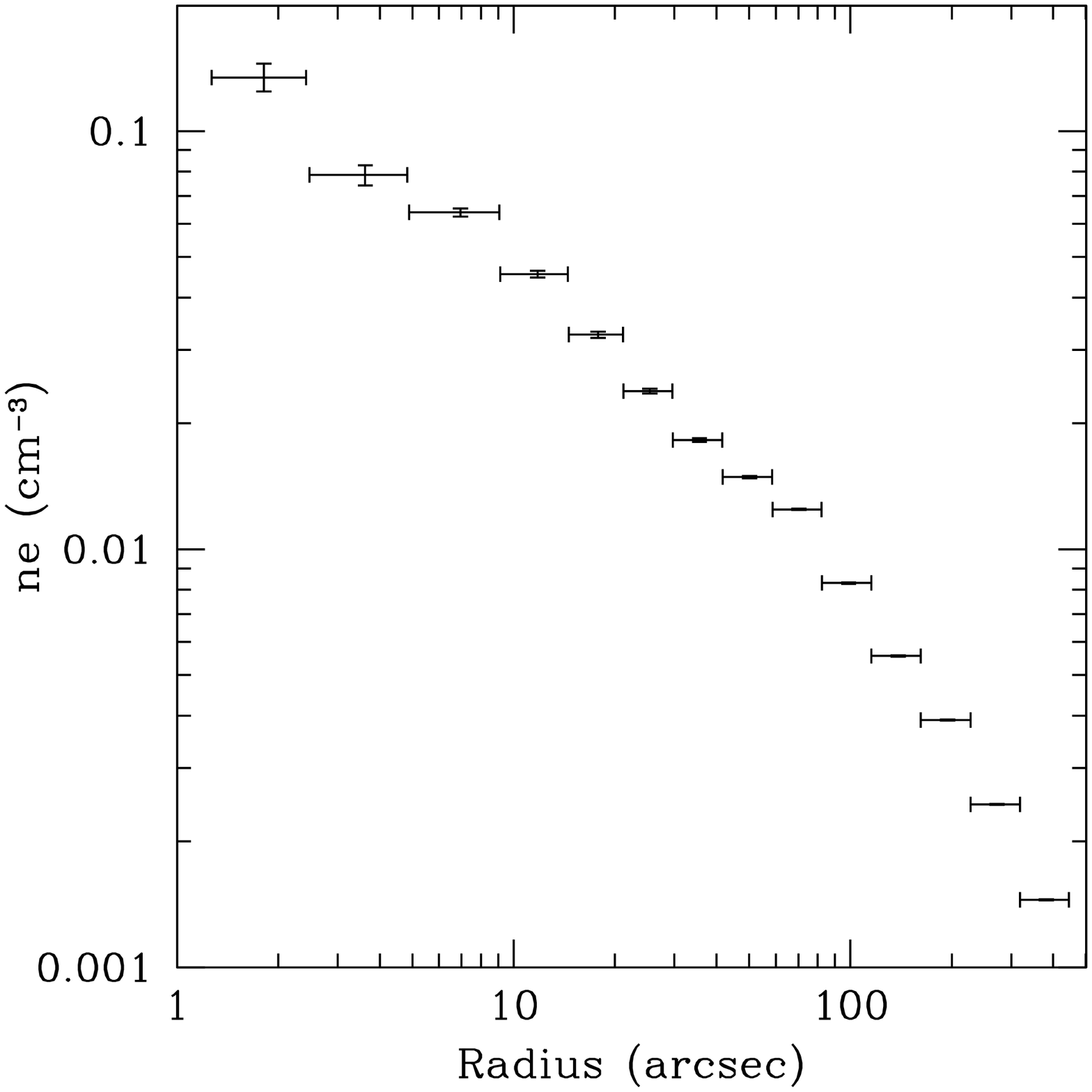,height=0.48\textwidth}
}
\caption{Azimuthally averaged temperature and density profiles.  {\sl
    Left:} Projected (red) and deprojected (black) temperature
  profiles.  {\sl Right:} Deprojected density profile.  Temperature
  error bars are 1 $\sigma$.}
\label{fig:deproj}
\end{figure*}

\subsection{Azimuthally Averaged Gas Temperature and Gas Density Profiles}

To obtain azimuthally averaged radial distributions of the gas
temperature, density and metallicity, spectra were extracted from
annuli centered on the active galactic nucleus (AGN) of NGC~6166.
These were fitted using a thermal plasma model \citep[\xspec{}
\apec{}, using the cosmic abundances of][]{ag89} with the Galactic
foreground absorption (\wabs, with fixed $N_{\rm H} = 9 \times 10^{19}
\rm\ cm^{-2}$).  The model was also combined with the \xspec{}
\projct{} model to obtain deprojected temperatures, densities and
abundances.  The resulting temperature and density profiles shown in
Fig.~\ref{fig:deproj}, with projected temperatures in red and
deprojected temperatures in black, are broadly consistent with
previous results \citep{jaf02, m11}.  The deprojection model assumes
spherical symmetry, which is clearly not accurate for A2199, resulting
in fluctuations in the deprojected temperature profile that are
unlikely to be real.  Nevertheless, the deprojected temperatures are
systematically lower than the projected temperatures, as usual, in the
cool core.  They rise from $\simeq 1.4$ keV in the cluster center to a
peak of $\simeq 5$ keV at a radius of $\simeq 200''$, beyond which
they decline slowly.  The density profile exhibits a break at $\simeq
100''$ associated with the SE edge.  There is also structure related
to the southwest excess (Fig.~\ref{fig:excess}), although this is
obscured by the coarse radial binning in Fig.~\ref{fig:deproj}.  The
metallicity declines from $\simeq 1.7$ solar at the cluster center to
$\simeq 0.3$ solar at a radius of $400''$, also in broad agreement
with previous results \citep{jaf02, m11}.

\subsection{Maps of Gas Properties}

\begin{figure*}
\centerline{
\epsfig{figure=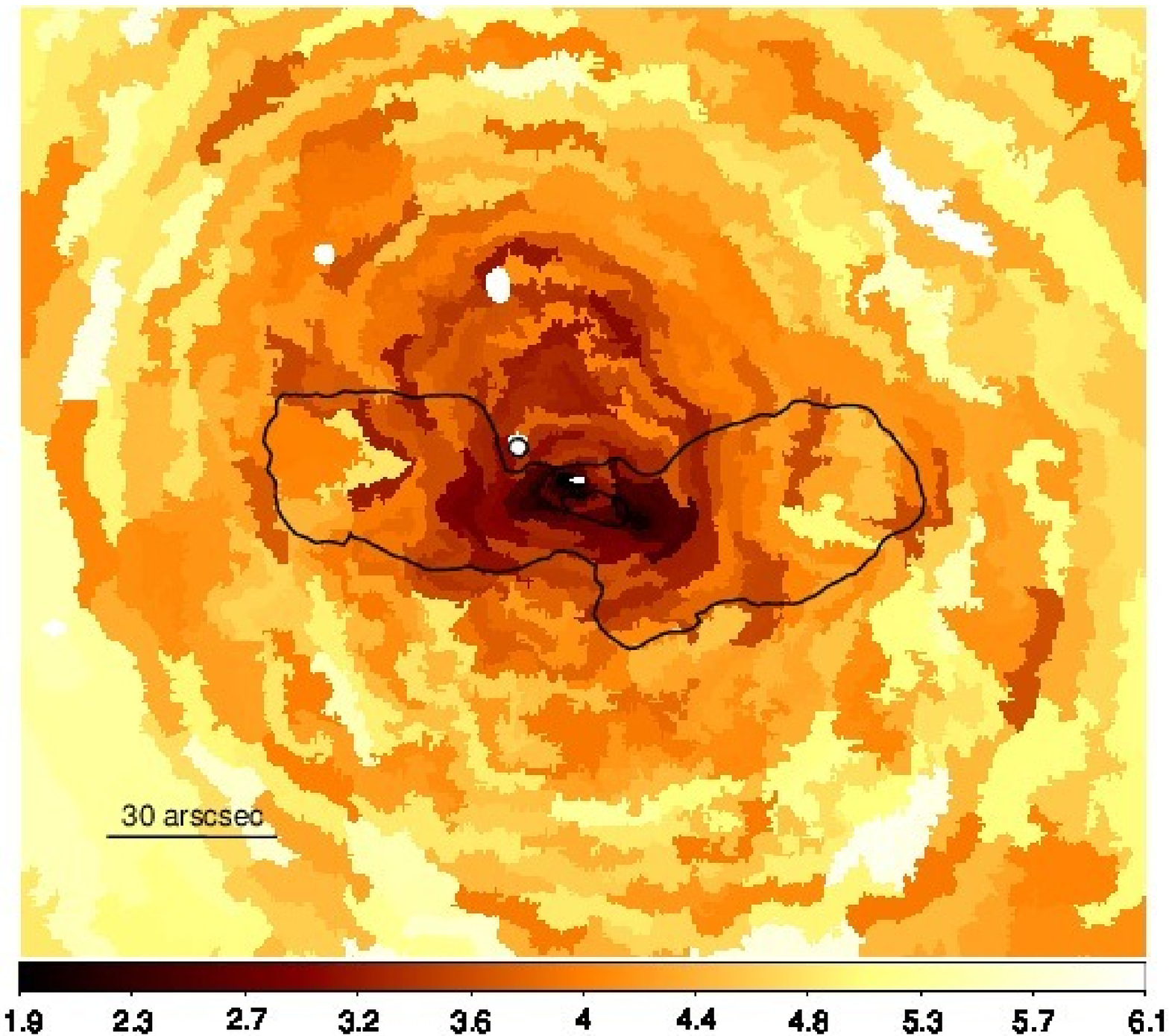,width=0.5\textwidth,clip=}
\epsfig{figure=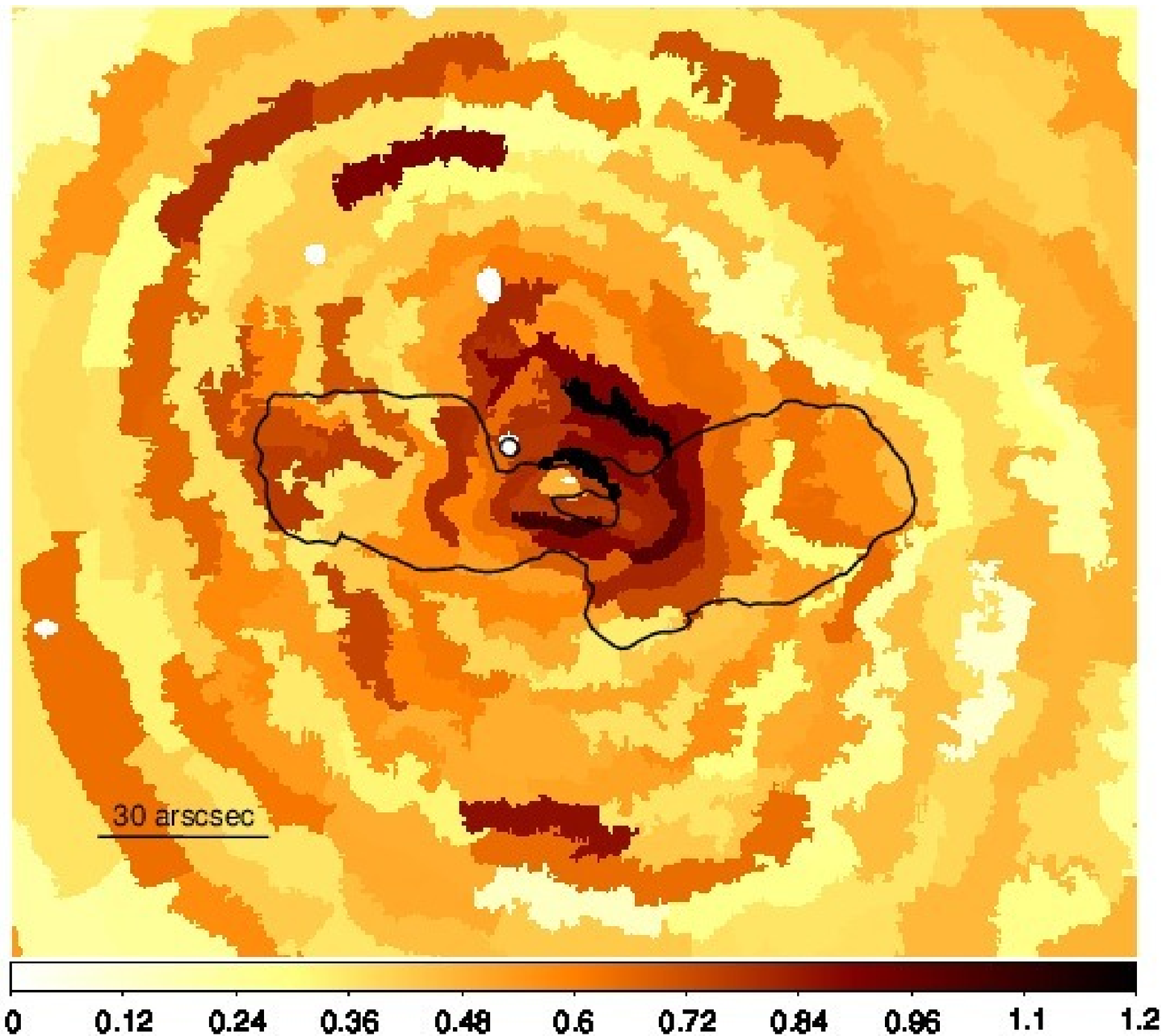,width=0.5\textwidth,clip=}
}
\caption{{\sl Left:} Temperature map of the central region of A2199.
  The color bar shows $kT$ in keV.  The black contour in both panels,
  which corresponds to a flux level of $2\times10^{-5}\rm\ Jy\
  beam^{-1}$ in the 5.9 GHz radio image (Fig.~\ref{fig:radio}), shows
  the extent of the radio lobes. {\sl Right:} Metallicity map for
  A2199.  The colour bar shows metallicity in units of the solar
  abundances of \citet{ag89}.  Deleted point sources appear white in
  both maps and the small white region at the center of each map marks
  the location of the AGN.}
\label{fig:tmap}
\end{figure*}

Maps of the gas temperature and metallicity were made using
reprocessed data for the four ACIS-I exposures (Table~\ref{tab:log}).
The data were binned into regions, defined using the contour binning
algorithm of \citet{s06} to obtain a specified signal to noise ratio
for each region.  The ratio between the length and width of each
region was constrained by setting the $C$ parameter of \citet{s06} to
2.  Spectra were summed and \xspec{} was used to fit them on the 0.5
-- 7 keV energy range by minimizing the C-statistic.  For the
temperature map, the target signal to noise ratio was 32 (1024 counts
per region).  An \apec{} thermal model, with fixed foreground
absorption and free metallicity, was fitted to each spectrum to obtain
the temperature map in the left panel of Fig.~\ref{fig:tmap}.  The
color bar there gives the temperature scale in keV.  Typical $1
\sigma$ errors range from about $\pm 5\%$ near the center of the map
to $\pm 10\%$ at the periphery.  For the metallicity map, the target
signal to noise ratio was 49 (2400 counts per region).  Fitting the
same thermal model as used to obtain the temperature map to the
spectra gave the metallicity map in the right panel of
Fig.~\ref{fig:tmap}.  The metallicity scale, which is given relative
to the solar abundances of \citet{ag89}, is shown in the color bar.
Typical uncertainties in this map are large, ranging from 18\% at the
center to 50\% at $200''$ from the center.  As discussed in section
\ref{sec:center}, improved spatial resolution combined with the
contour binning reveal structure in the temperature and metallicity
maps that was not evident in the previous maps of \citet{jaf02}.

\subsection{VLA Radio Structures}

\begin{figure*}
\centerline{
\epsfig{figure=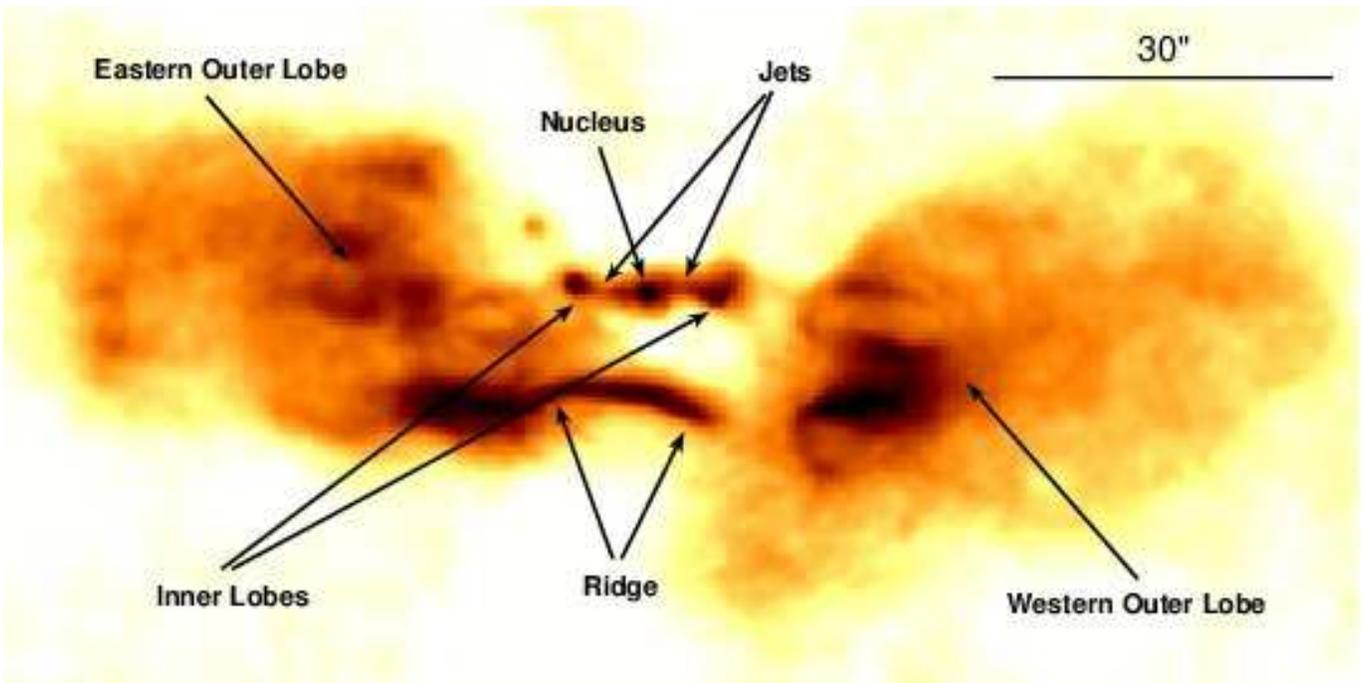,width=\textwidth,clip=}
}
\caption{4.9 GHz VLA flux density map of 3C\,338. Prominent radio
  structures are labeled including inner and outer radio lobes, a
  two-sided jet, and a ridge of radio emission lying to the south of
  the nucleus and extending east-west to the outer radio lobes.
  The color scale is logarithmic in the radio flux.
}
\label{fig:radio}
\end{figure*}

Like many other clusters, groups, and early-type galaxies, the X-ray
emitting gas in A2199 shows evidence of outbursts from an AGN located
at the peak of its X-ray emission.  The radio source 3C~338, which is
hosted by NGC~6166, provides clear evidence of outbursts powered by
accretion onto a supermassive black hole in the nucleus of NGC~6166.
Structures associated with 3C~338 include \citep[Fig.~\ref{fig:radio};
as discussed by][]{bsw83, grt07}:
\begin{itemize}
\item  a bright nucleus
\item a two-sided jet with an extent of $\sim$15$^{\prime\prime}$
\item a pair of inner radio lobes ($\sim7''$ from the nucleus) at the
  ends of the two-sided jet 
\item two outer radio lobes centered $\sim$40$''$ from the nucleus
\item an unusual ``ridge'' lying $\sim10''$ south of the nucleus,
  first reported by \citet{bsw83}.
\end{itemize}

The jet is a region of diffuse radio emission linking the nucleus to
the two spatially extended inner radio lobes.  While the inner lobes
are aligned with the outer radio lobes, there is no convincing
connection between them in the radio.  The radio ridge runs roughly
parallel to the jet and appears to connect to the two outer lobes.
\citet{bsw83} argued that the ridge was the result of a dynamical
interaction with the ICM.  Detailed VLBI as well as VLA radio studies
of 3C\,338 also are presented by \citet{fcg93} and \citet{grt07}.
Equipartition properties for the inner radio lobes (cavities) and the
ridge are given in Table~\ref{tab:radio2}.  Note that \citet{bsw83}
estimated the minimum pressure of the ridge using $2 B_{\rm min}^2 /
(8 \pi)$, whereas here we use $(7/9) \times B_{\rm min}^2 / (8 \pi)$,
consistent with the minimum energy condition and a magnetic field that
is not well ordered.  This accounts for the greater part of the factor
of 5 difference between their minimum pressure estimate and ours.

\subsection{Core Multi-wavelength Structures} {\label{sec:center}}

At smaller angular scales, interior to the SE edge ($<100''$), there
is a wealth of substructure apparent in multi-wavelength maps of
A2199.  To show the relationship between the radio and X-ray features,
radio contours are plotted on the residual image in the right panel of
Figure~\ref{fig:central}.  The ICM appears highly disturbed in this
region.  Structures include:
\begin{itemize}
\item two ``outer cavities'' to the east and west of the cluster
  center, decrements in the X-ray emission that coincide with radio
  lobes (\S\ref{sec:shock}), also discussed by \citet{jaf02} 
\item a filament (``Shell'') partially enclosing the western X-ray
  cavity
\item a third X-ray depression (``SW Cavity'') coincident with a spur
  of radio emission extending southward from the western outer lobe,
  first noted by \citet{grt07}
\item a plume of emission extending about $50''$ north from the bright
  core 
\end{itemize}

The ``shell'' of enhanced X-ray emission along the periphery of the
western outer lobe and hints of a similar feature around the eastern
outer lobe can be seen in Figs.~\ref{fig:fov} and \ref{fig:central}.
While these features are irregular and not well defined, they do
correspond to regions of reduced temperature in the map of
Fig.~\ref{fig:tmap}.

\begin{figure}
\centerline{
\epsfig{figure=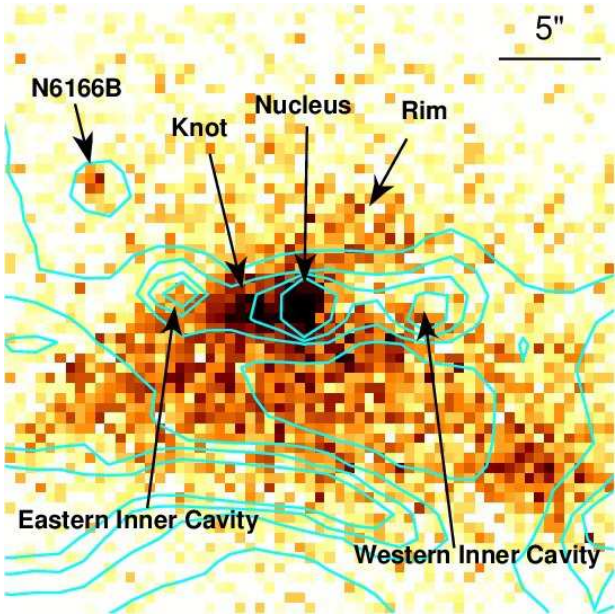,width=0.45\textwidth,angle=0,clip=}
}
\caption{0.5 -- 2 keV image of the central $30'' \times 30''$, with
  several X-ray features labeled.  Contours of the 4.9 GHz radio
  emission are overlaid.  The color scale is logarithmic in the
  surface brightness.}
\label{fig:core}
\end{figure}

Within the central $30^{\prime\prime}\times30^{\prime\prime}$
(Fig.~\ref{fig:core}), the region with the brightest X-ray emission is
roughly triangular in shape, bounded by the jet to the north and by
the inner edges of the two outer radio lobes to the east and west,
giving it an east-west extent of $\simeq 30''$
(Fig.~\ref{fig:central}).  The radio ridge lies over this region,
roughly matching its southward curvature.  Gas in the region is cool
(Figs.~\ref{fig:tmap} and \ref{fig:mina}).  An X-ray deficit coincides
with the western inner radio jet, reminiscent of a ``tunnel'' along
the jet path \citep{csb05}.  An X-ray bright filament or ``rim'' lies
along the northern edge of the western jet.  Along the eastern jet,
there is no deficit in the X-ray emission; instead we find a ``knot''
of enhanced emission located $\sim 2''$ from the X-ray/radio nucleus
(0.5 -- 8 keV luminosity $\simeq 4 \times 10^{40} \rm\ erg\ s^{-1}$;
see section \ref{sec:bright_spec}). There is marginal evidence of an
X-ray cavity (``eastern inner cavity'') east of the knot, coinciding
with the eastern inner radio lobe and bounded by enhanced X-ray
emission in the south.

\subsubsection{Optical Line-emitting Gas}

\begin{figure}
\centerline{
\epsfig{figure=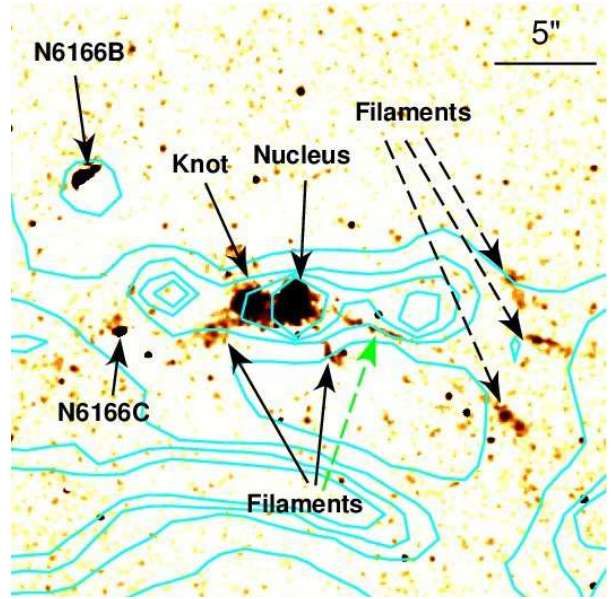,width=0.45\textwidth,bbllx=15,bblly=15,
  bburx=556,bbury=554,clip=}
}
\caption{An H$\alpha + \rm[N\ II]$ image of the core region of A2199,
  smoothed by a Gaussian kernel of $\sigma = 0.2''$. The image is
  obtained by subtracting an {\sl HST}/ACS F814W (continuum) image
  from an FR656N ($\rm line + continuum$) image.  Several
  H$\alpha$-emitting features, including the two galaxies (NGC~6166B
  and NGC~6166C) in projection \citep{l86}, are labeled.  VLA 4.9 GHz
  flux density contours are overlaid. The H$\alpha$ filaments marked
  with dashed arrows appear related to the propagation of the radio
  jet.  The color scale is logarithmic in image brightness.}
\label{fig:Halpha}
\end{figure}

Cooler gas in the core region of A2199 manifests itself in optical
emission lines and dust extinction \citep{mfb04}.  We used archival
{\sl HST} data to reproduce the H$\alpha + \rm [N\ II]$ image
presented in \citet[Fig.~3 therein]{mfb04} and discussed by them in
detail.  As shown in Fig.~\ref{fig:Halpha}, there is clumpy H$\alpha$
emission at the positions of the nucleus and the X-ray knot, with an
extent of $\sim2''$ in both cases.  Several H$\alpha$ filaments can
also be seen within $\sim 10''$ of the nucleus. In particular, one
filament (marked by a green dashed arrow in Fig.~\ref{fig:Halpha})
appears to follow the southern edge of the western jet.  Another three
filaments (marked by black dashed arrows in Fig.~\ref{fig:Halpha}) are
located immediately west of the western inner radio lobe, indicating a
connection between the inner and outer lobes, although the radio
emission itself does not unambiguously show such a connection.  We
have measured the H$\alpha + \rm [N\ II]$ luminosity and estimated the
corresponding densities and masses of ionized gas for several distinct
features, including the nucleus, the knot and the filament apparently
following the southern edge of the western jet.  Like \citet{mfb04},
we adopt a [N II]/H$\alpha$ line ratio of 2, typical of narrow line
regions around active nuclei.  The results and details of the regions
are summarized in Table~\ref{tab:Halpha}.

The H$\alpha$ filaments associated with the inner edge of the
western outer radio lobe may be entrained and drawn outward
in the flow associated with the radio lobes.  They resemble similar
features in the line emitting gas seen in the Perseus cluster
\citep{fsc03}.

\subsubsection{X-ray Spectra of Bright Regions} \label{sec:bright_spec}

\begin{figure*}
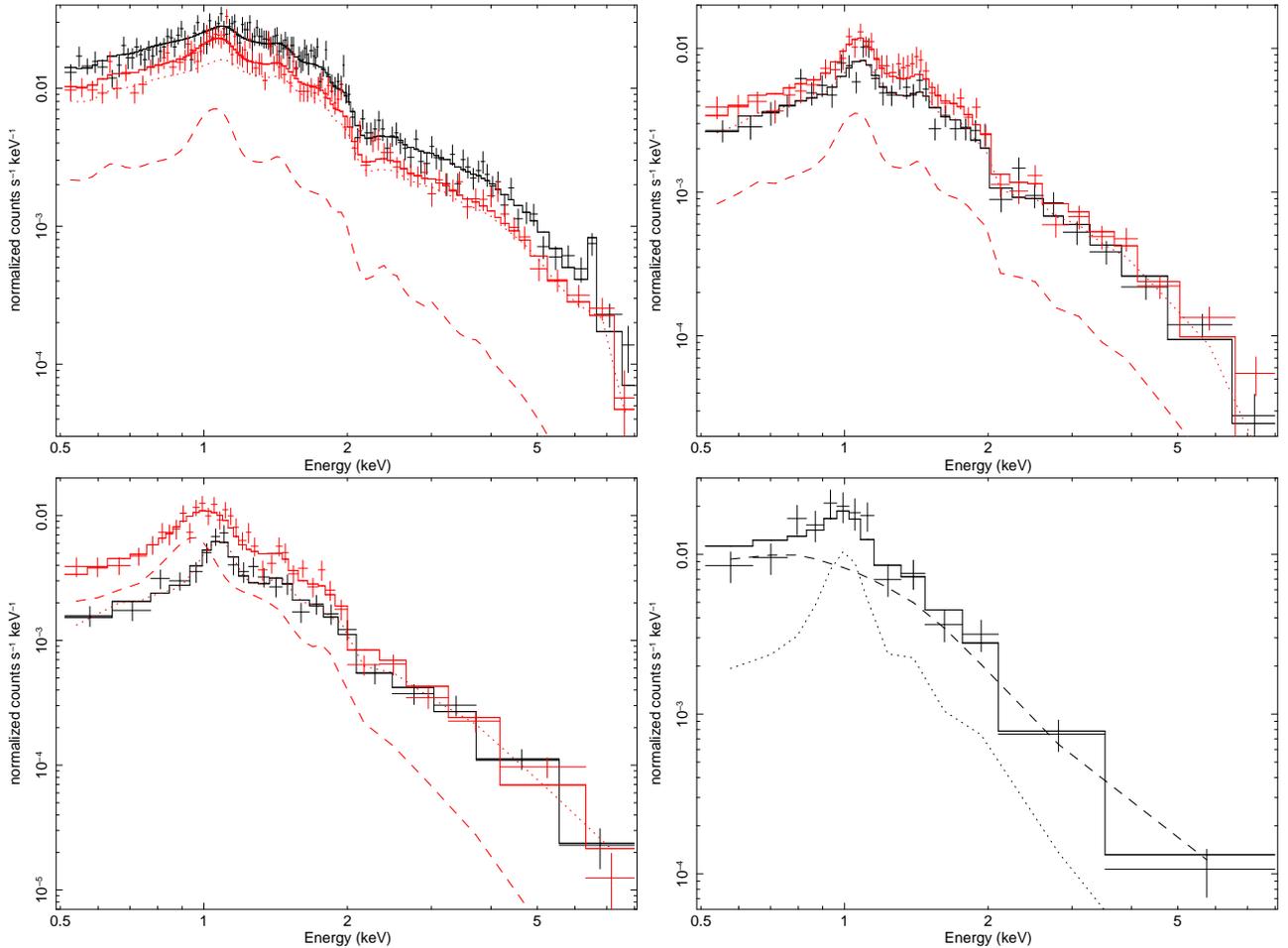

\centerline{
\epsfig{figure=fig10a.ps,width=0.35\textwidth,angle=270,clip=}
\epsfig{figure=fig10b.ps,width=0.35\textwidth,angle=270,clip=}
}
\centerline{
\epsfig{figure=fig10c.ps,width=0.35\textwidth,angle=270,clip=}
\epsfig{figure=fig10d.ps,width=0.35\textwidth,angle=270,clip=}
}

\caption{Fitted spectra for (left to right, top to bottom): (a) the
  shell; (b) the rim; (c) the knot; (d) the nucleus.  In (a), (b), and
  (c), the red and black spectra are the source and local background
  spectra, respectively.  Best-fit models are shown by the solid
  curves.  For the background spectra, a single temperature model is
  adopted; the source model includes an additional thermal model
  (dashed curves).  In (d), the nuclear spectrum from ObsID 497 is
  fitted with a power-law (dashed curve) plus an \apec{} (dotted curve)
  model.}
\label{fig:spec}
\end{figure*}

We extracted X-ray spectra from the ``shell'', the ``rim'', and the
``knot''.  All three features either enclose or overlie a volume
filled with energetic particles.  For each feature, a local background
region, immediately adjacent to the source region, was fitted with a
thermal model that was scaled, with other parameters fixed, and used
to account for emission by overlying gas (region details in
Table~\ref{tab:spec} notes).  The excess emission in the source region
was then modeled by an additional thermal component.  The fit results
are summarized in Table~\ref{tab:spec} and the spectra are shown in
Fig.~\ref{fig:spec}.  Use of a local background helps to separate the
emission of a brighter region from that of surrounding gas.  As a
result, temperatures in Table~\ref{tab:spec}, notably for the shell,
are lower than those at the same location in the temperature map.

The estimated density for the ``shell'' is inversely proportional to
the square root of its estimated volume, which is a significant source
of uncertainty.  Nevertheless, the shell is clearly denser than the
surrounding gas, since it stands out in the X-ray image
(Fig.~\ref{fig:central}).  It is also cooler than other gas
surrounding the outer end of the western radio lobe (for which $kT =
3$ -- 4 keV; Fig.~\ref{fig:deproj}).  Together, these properties show
that the shell has lower entropy than the surrounding gas.  From
Table~\ref{tab:spec}, the entropy index of the ``shell'' is 
$\Sigma = kT / n^{2/3} \simeq 16 \rm\ keV\ cm^{-2}$, whereas $\Sigma
\simeq 60 \rm\ keV\ cm^{-2}$ for the surrounding gas based on the
deprojected temperature and density profiles at $r \simeq 50''$
(Fig.~\ref{fig:deproj},).   As in other cool core clusters hosting
radio outbursts, this low entropy has probably been lifted outward
with the radio lobes \citep[e.g.,][]{fjc07, gnd11}.  Gas with
comparable entropy is found at $\simeq 10''$ from the cluster center
(Fig.~\ref{fig:deproj}), suggesting that the gas in the shell
originated from within about this radius.

The spectra of the X-ray ``knot'' in the eastern radio jet are well
fitted by a thermal plasma model with a temperature of 1.0 keV; a
power-law model is formally excluded ($\chi^2/\rm d.o.f. = 77/35$).
Compact X-ray features are seen in the jets of a number of
Fanaroff-Riley class I radio galaxies \citep[FR I;][]{fr74}, but these
are generally interpreted as synchrotron emission from
ultra-relativistic particles created by internal shocks in the jet
\citep{wb06}.  Also, while the estimated pressure in the knot is
roughly twice that for the innermost shell of the deprojection, it is
similar to the estimated pressure for the ``rim,'' so that it is not
greatly overpressured compared to other gas at the same radius.  The
knot's thermal spectrum, its H$\alpha$ counterpart, the lack of a
radio counterpart, and its similar pressure to other gas features at
the same radius all argue that the knot is a feature of the ICM and
not internal to the jet.  At the least, it is not greatly affected by
the jet.  It might be gas stripped from one of the massive early-type
galaxies that lie near the center of the cluster.

Although it lies adjacent to the jet, the gas in the ``rim'' is cooler
than gas lying immediately to its north, further from the jet.  This
largely rules out the possibility that the bright rim consists of gas
that has been shocked during recent activity of the radio source.  It
is more likely to be gas lifted outward in the flow associated with
the jets and radio lobes.

\subsubsection{Northern Plume} \label{sec:plume}

\begin{figure}
\centerline{
\epsfig{figure=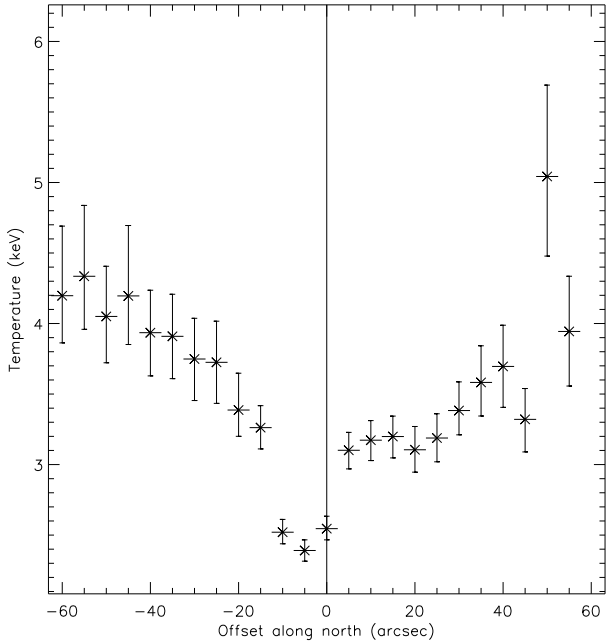,width=0.5\textwidth,angle=90,clip=}
}
\caption{Distribution of gas temperatures along a north-south axis
  through the cluster center.  Northern offsets are positive and the
  location of the AGN is marked by a vertical line.  Apart from
  the cool region extending $\simeq 10''$ south of the AGN, gas in the
  plume that extends $\simeq50''$ north of the AGN is systematically
  cooler than gas to the south.  Both the plume and the cool gas to
  the south are consistent with sloshing.}
\label{fig:mina}
\end{figure}

Finally, a ``Northern Plume'' of enhanced emission extends $\sim 50''$
north of the nucleus.  This feature corresponds to a cooler region in
the temperature map of Fig.~\ref{fig:tmap}.  It is less evident due to
the larger errors in the abundance map, but still present as a region
of enhanced abundances (Fig.~\ref{fig:tmap} right).  We extracted
spectra for a set of adjacent slices along an axis from south to north
through the AGN.  The east-west extent of each slice is $30''$ and
their widths are $5''$.  Each spectrum was fitted with an absorbed
\apec{} model.  The fitted temperature distribution is shown in
Figure~\ref{fig:mina}.  While the X-ray brightest regions in the core
between the jet and the radio ridge (Fig.~\ref{fig:core}) have the
lowest temperatures of $\simeq 2.5$ keV, temperatures in the northern
plume are systematically lower than temperatures at the same distance
to the south of the center, by as much as 0.5 keV.  \citet{jaf02}
suggested that the northern plume is a cooling wake left behind the
moving core.  However, the cooling time of gas 40'' north of the AGN
comfortably exceeds 1 Gyr, whereas the free-fall time from there to
the AGN is less than 100 Myr.  Thus, the transient impact of the
moving cD occurs on a timescale that is an order of magnitude shorter
than the gas cooling time, making its effect on gas cooling minor.  By
contrast, the cooling time of the gas within 3'' of the AGN
(Fig.~\ref{fig:deproj}) is an order of magnitude shorter than that of
the gas at 40'', making the inner gas far more prone to cooling.  This
is difficult to reconcile with a cooling wake and we do not consider
that possibility any further.  In section \ref{sec:slosh} we suggest
this feature is due to sloshing of the core gas \citep{mvf03}.

\subsection{The Nucleus}

The X-ray nucleus and the underlying accretion onto the central
supermassive black hole of NGC~6166 have been studied by
\citet{dja01}.  Our deep observations allow us to better constrain the
spectral properties of the nucleus, as well as the circumnuclear gas.
We extracted spectra from the central $1\farcs5$ for individual
observations.  A model consisting of a power-law for the AGN plus an
\apec{} thermal spectrum for the emission from the circumnuclear hot
gas was used to fit the spectra.  Since no statistically significant
flux or spectral variation was found between the observations, all six
spectra were fitted jointly, giving an acceptable fit with a
photon-index of $2.25^{+0.36}_{-0.19}$, a gas temperature of
$1.32^{+0.55}_{-0.31}$ keV, and an absorption column of
$9.9^{+7.6}_{-4.4}\times10^{20}{\rm~cm^{-2}}$.  We estimate an 0.5 --
8 keV intrinsic luminosity of $1.2_{-0.4}^{+0.3} \times 10^{41}
{\rm~erg~s^{-1}}$ for the nucleus, and a density of
$0.35^{+0.19}_{-0.09}{\rm~cm^{-3}}$ for the circumnuclear gas.  The
cooling time of this gas is very short, at $\simeq 26$ Myr, supporting
the case for AGN feedback in A2199.  However, since it is more than an
order of magnitude longer than the dynamical time, the gas can remain
hydrostatic and we should not expect catastrophic cooling.

\section{The Physical Origin of X-ray and Radio Structures in A2199 --
  Gas Sloshing and AGN Outbursts} {\label{sec:discussion}}

Here we discuss the origin of several features as the products of
large scale sloshing and outbursts from the AGN in NGC~6166.

\subsection{Large Scale Gas Sloshing} \label{sec:slosh}

The most prominent large scale feature is excess emission to the
southwest and west of the cluster center (Figs.~\ref{fig:excess} and
\ref{fig:profiles}).  The most plausible cause of asymmetries on such
a large scale, $\sim100$ kpc, is merging activity.  In the absence of
an obvious remnant gas core, these features are most likely to be the
response of the cluster gas to an infalling subcluster, as found in
many other clusters and demonstrated by simulations \citep[see review
of][]{mv07}.  We refer to such effects as sloshing.

Although on a considerably smaller scale, the other feature that is
strongly suggestive of gas sloshing is the dense, cool, northern plume
(Figs.~\ref{fig:central} \& \ref{fig:mina}).  Simulations show that
sloshing leads to the formation of contact discontinuities in the gas,
i.e., cold fronts, where the pressure is continuous, but there is an
abrupt change in the density \citep[e.g.,][]{am06, zmj10, rbs11}.  The
cold fronts are embedded in a larger scale, spiral shaped density
enhancement that can extend much further than the fronts from the center
of the perturbed cluster.  The northern plume is roughly consistent
with such a spiral density enhancement, seen approximately edge on.
Its orientation and scale suggest that it could well be part of the
same disturbance as the southwest excess.  If so, we are viewing A2199
from close to the plane of the orbit of the perturber.  In this
geometry, the temperature and brightness profiles of the front are
sensitive to the viewing direction.  Nevertheless, the temperature and
brightness of the plume are broadly consistent with sloshing simulations.

The plume ends abruptly $\sim
45'' \simeq 28$ kpc north of the cluster center.  \citet{rbs11} find
that the outermost sloshing cold front moves outward at a nearly
constant speed of $\simeq s/13$, where $s$ is the local sound speed.
If the outer edge of the northern plume is the outermost sloshing cold
front, then it can provide an estimate of the time since the perturber
passed through the core of A2199.  For a distance of 28 kpc and a
local gas temperature of 3.3 keV, the age would be $\simeq
4\times10^8$ yr.  While the lopsided appearance of the northern plume
supports its interpretation as a sloshing cold front, since we are
viewing the sloshing spiral nearly edge on, we cannot tell if the edge
of the plume is the outermost cold front.  Despite this,
based on the simulations, the unknown
geometry contributes no more than 50\% uncertainty, comparable to the
systematic uncertainty in our estimate for the speed of the front
(which is based on a limited set of simulations).  This age estimate
exceeds those for the radio outburst that produced the $100''$ shock
by an order of magnitude (see section \ref{sec:shock}), so
we can be confident that the shock is more recent.

\begin{figure*}
\centerline{
\epsfig{figure=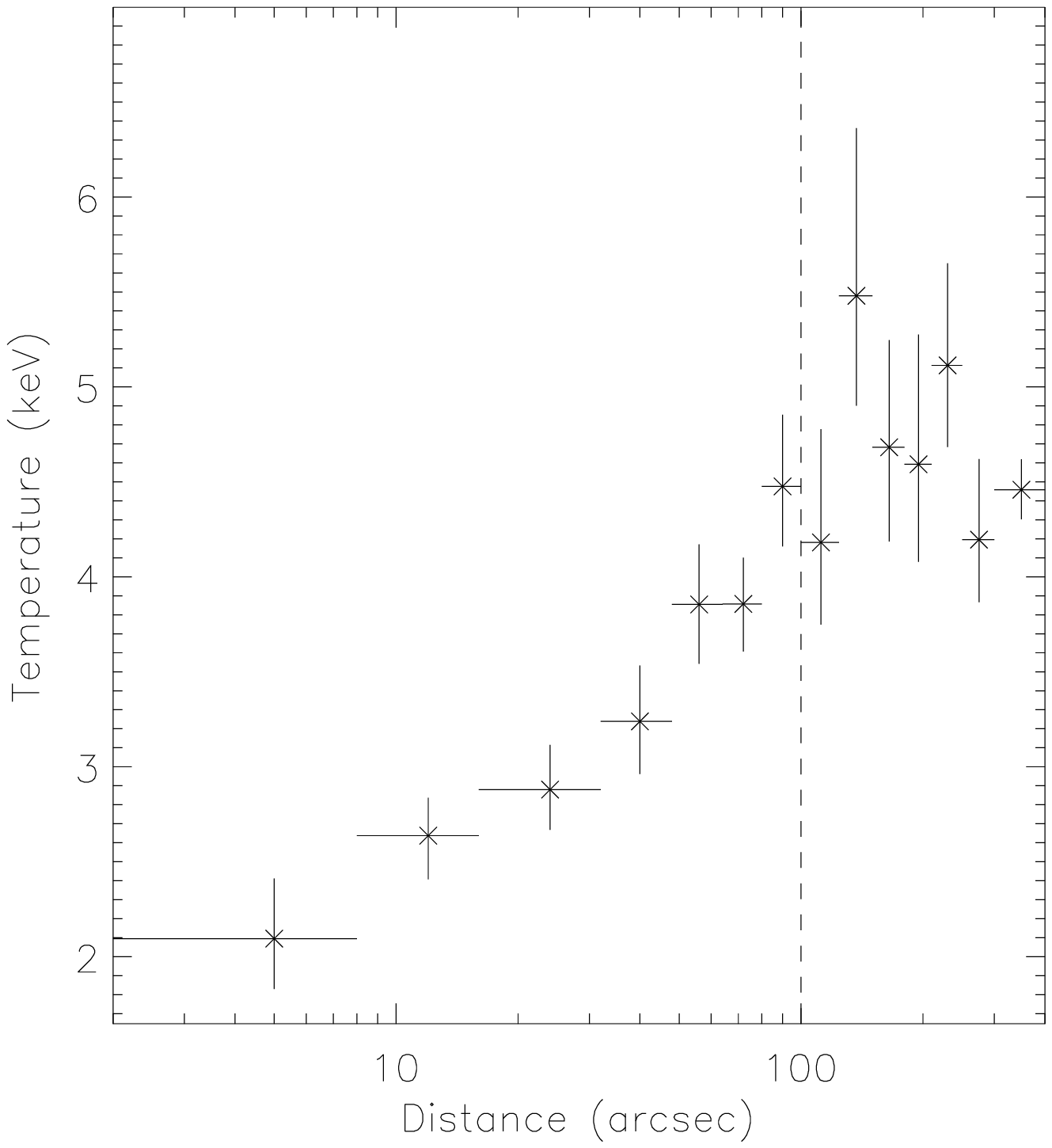,width=0.5\textwidth,angle=90,clip=}
\epsfig{figure=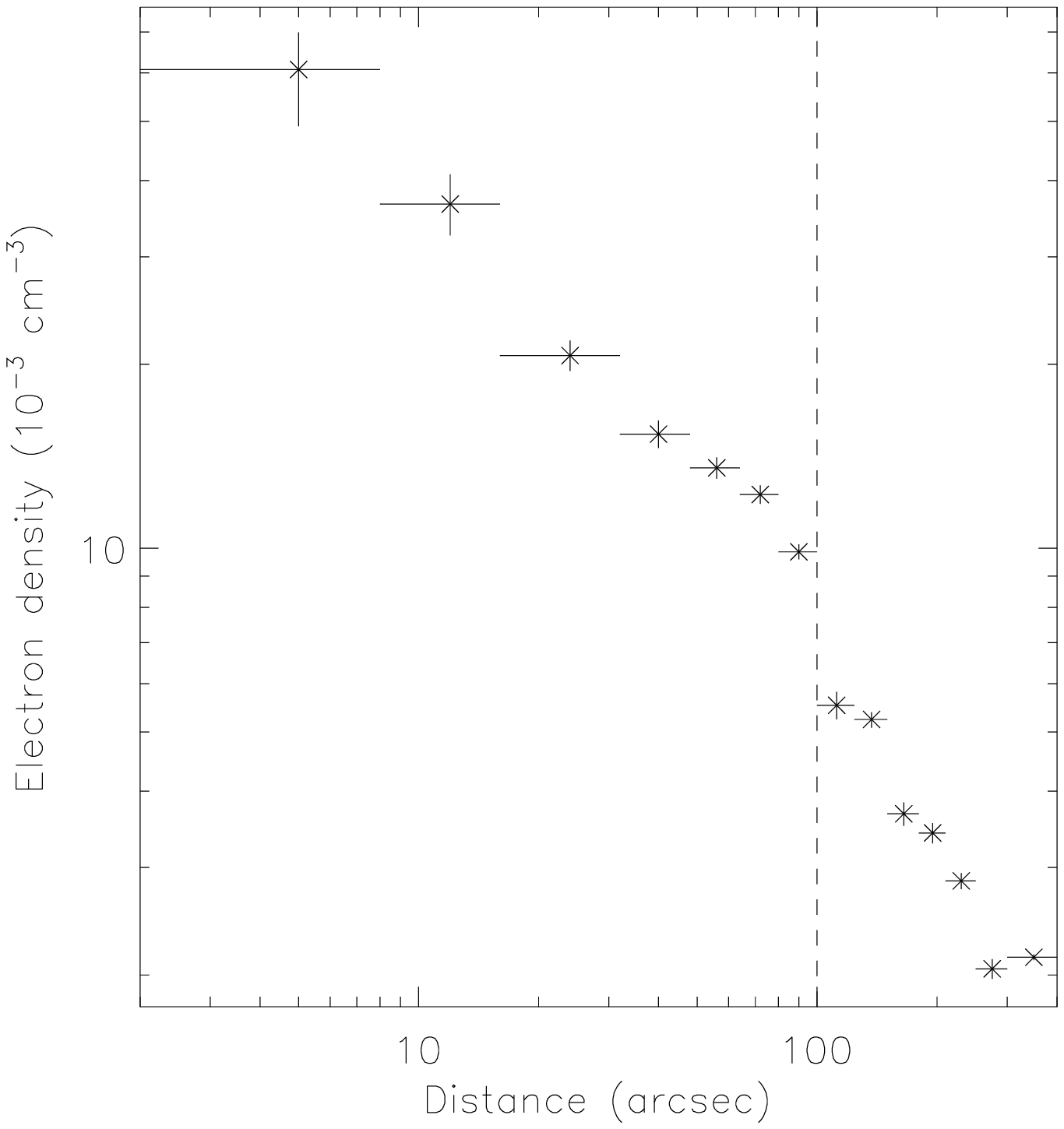,width=0.5\textwidth,angle=90,clip=}
}
\caption{Deprojected temperature (left) and density (right) across the
  SE edge.  The position of the surface brightness discontinuity is
  shown by the dashed vertical line. The error bars are of 1 $\sigma$.
  The gas density shows a prominent edge at $100''$ while there is no
  clear change in the gas temperature.}
\label{fig:front_deproj}
\end{figure*}

\subsubsection{Sloshing and the Radio Ridge} \label{sec:ridge}

The morphology of the radio ridge is very suggestive of an old radio
jet that has become detached from the AGN in NGC~6166
\citep[Fig.~\ref{fig:radio};][]{bsw83}.  This could be a result of
sloshing.  Since the radio plasma is effectively weightless, in the
presence of sloshing, it is expected to move with the ICM.  The gas
within about 10'' south of the AGN is brighter (Fig.~\ref{fig:core}),
so clearly denser, and also cooler (Fig.~\ref{fig:mina}) than the gas
to the north, so it has lower entropy.  In equilibrium, the lowest
entropy gas should be centered on the AGN.  The southward displacement
of the low entropy gas on much the same scale as the radio ridge is
consistent with the ridge simply moving along with the gas
(Fig.~\ref{fig:core}).  It is remarkable that the ``jet'' could
survive being swept away from the AGN largely intact, which requires
the gas flow to be laminar.  \citet{bsw83} have discussed some other
challenges for the interpretation of this feature as the remnant of a
jet.  One such constraint is the synchrotron lifetime of the electrons
radiating at 5 GHz, which is $\simeq 7.5$ Myr for the
``equipartition'' magnetic field of $18\ \mu\rm G$
(Table~\ref{tab:radio2}).  Since the projected separation of the ridge
and the AGN is $\simeq 6$ kpc, the mean speed of the ridge relative to
the AGN would need to exceed $\simeq 800\rm\ km\ s^{-1}$ for the
emitting electrons to survive the trip.  While this speed is
implausibly fast for sloshing, speeds a factor of $\sim 2$ smaller are
more reasonable.  Furthermore, motion of the galaxy excited along with
the sloshing could boost the speed of the AGN relative to the gas
\citep{am06}.  The magnetic field in the ridge may also be small
enough to be consistent with a lower speed.  Typical of FR I radio
sources, the equipartition pressures for the radio features given in
Table~\ref{tab:radio2} are more than an order of magnitude smaller
than the pressure in the ICM at $\sim 7$ kpc from the AGN, leaving
latitude for substantial departures from equipartition and allowing
magnetic field strengths significantly smaller than $18\ \mu\rm G$ .

\subsection{Southeastern Edge}

The next prominent feature interior to the large scale asymmetry is
the southeastern (SE) edge, seen in Figs.~\ref{fig:fov} and
\ref{fig:excess} at a distance of $\sim100''$ from the center of A2199
and discussed by \citet{jaf02} and \citet{sf06}.
Fig.~\ref{fig:central} shows an enlarged view of the core.  To
understand the origin of the SE edge, we extracted spectra from a set
of sectors across the edge, which span an azimuthal range of
30$^\circ$-180$^\circ$ (east from north).  The \xspec{} \projct{}
model was used to derive the density and temperature profiles
(Fig.~\ref{fig:front_deproj}).  Across the SE edge, the gas density
decreases by a factor of $\sim 1.7$, whereas the gas temperature
exhibits no significant change, suggesting a pressure ratio across the
front of $\sim 1.7$.  These values rely on assuming that the gas
distribution is spherically symmetric, a source of systematic error
that could affect the following discussion.  We consider three
possible origins for this feature --- 1) a classical cold front
created by the remnant core from a merger, 2) a sloshing cold front
and 3) a shock front.  As discussed below, although we prefer the
shock interpretation, both a sloshing cold front and a shock front
could be consistent with the observations and have some supporting
evidence.

\subsubsection{Remnant Core Cold Front}

A remnant core cold front \citep[e.g., like that in A3667,][]{vmm01}
occurs at the contact discontinuity between low entropy gas from the
remnant core of a merging cluster or group and the ICM.  This would
account for the higher gas density interior to the front.  However, at
a simple contact discontinuity we should expect to find pressure
equilibrium, which would require the gas temperature to be a factor of
$\simeq 1.7$ higher on the low density side of the front, which it
clearly is not.  If we disregard the first data point beyond $100''$
in Fig.~\ref{fig:front_deproj}, it might be argued that the
temperature is declining outside the front, in which case the
temperature at the front might be as high as $\sim 5.5$ keV.
Nevertheless, it is difficult to argue that the temperature on the
outside of the front is high enough ($\gtrsim6.8$ keV) to make the
pressures equal on either side of the front.  Certainly, the data are
not easy to reconcile with the pressure being continuous at this
front.

There are other issues for such a model.  The SE edge does not
resemble other well known examples of remnant core cold fronts.  For
example, the front in A3667 subtends an angle at the center of the
remnant core of no more than about $90^\circ$ (compared to more like
$180^\circ$ for A2199), beyond which the low entropy gas is truncated
\citep{ocn09}.  The front associated with the higher speed remnant
core of the Bullet cluster, 1E0657-558, has a well-defined head and a
flared tail, nothing like the nearly spherical front here \citep{m06}.
In addition, the core of a merging group or cluster moving through
A2199 would be expected to dramatically alter the fine radio structure
seen in Fig.~\ref{fig:radio}.  The outer (and inner) radio lobes
appear relatively undisturbed and roughly symmetrical.  It is
difficult to see how the transonic or supersonic passage of a core,
the size of the observed front, could be consistent with the observed
radio features. To be so little disturbed by the associated gas flows,
the lobes would need to have expanded recently and supersonically, in
which case we should expect to find prominent shocks closely
enveloping them.

\subsubsection{Sloshing Cold Front}

A sloshing cold front appears more viable than a remnant core cold
front, but is also problematic.  Like remnant core cold fronts, the
pressure should be continuous across sloshing cold fronts \citep[e.g.,
for A1795,][]{mvm01}.  As argued in the preceding section, the
pressure might be continuous if the temperature is rising inward
beyond the front (as one approaches the front from the east), which
might allow the interpretation as a cold front.  While it would be
unusual to have the temperature rising inward at such small radii,
this could be a consequence of gas sloshing.  However, the first
temperature point beyond the edge in Fig.~\ref{fig:front_deproj}
remains a problem for this interpretation with a simple geometry.

\begin{figure}
\centerline{\epsfig{figure=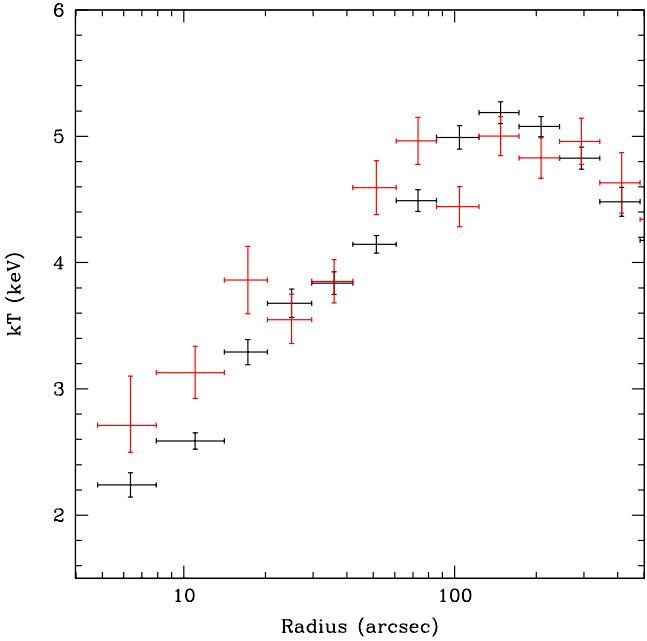,width=0.5\textwidth,angle=0}}
\caption{Temperature profiles for southeast and northwest sectors.
  Projected temperature is plotted against the radius for the sectors
  $45^\circ$ to $180^\circ$ (black) and $-67^\circ$ to $-28^\circ$
  (red).  Temperature error bars are $1 \sigma$.
} \label{fig:ktsectors}
\end{figure}

Figure~\ref{fig:ktsectors} compares profiles of the projected gas
temperature, in black for the southeast sector, in the azimuthal range
$45^\circ$ to $180^\circ$, where the SE edge is most prominent, and in
red for the northwest sector, $-67^\circ$ to $-28^\circ$, where the
edge at $100''$ is least evident.  The relatively lower temperatures
from $40''$ -- $80''$ in the southeast sector compared to the
northwest sector are broadly consistent with the alternating patterns
of entropy and temperature produced by sloshing \citep{am06, zmj10,
  rbs11}.  Set against this, the $100''$ edge is abnormally
round and extends over at least $180^\circ$ in azimuth
(Fig.~\ref{fig:central}), unlike other observed or simulated sloshing
fronts.

As discussed above (Fig.~\ref{fig:central}, \S\ref{sec:slosh}, and
\S\ref{sec:ridge}), the northern ``plume'' and radio ridge are
consistent with significant gas sloshing in the core region ($\lesssim
100''$), particularly when coupled with the larger scale southwest
excess.  The SE edge is roughly twice as far from the cluster center
as the end of the northern plume.  While an aspherical potential could
distort sloshing cold fronts away from a common plane, the center of
the SE edge is roughly $90^\circ$ away from the axis defined by the
southwestern enhancement and the northern plume, making it improbable
that these features were generated by the same perturbation.  Thus, if
the SE edge is the result of sloshing, there would need to have been
multiple merger events.  This raises no immediate issues.  In fact,
although they are unlikely to be the product of either merger event,
the well-known multiple nuclei of NGC~6166 \citep[including NGC~6166A,
B, C and D;][]{l86}, seen in projection within the central 50 kpc, are
consistent with a high rate of infall in this system.  Of course, for
a merger to cause the SE edge, it would be further in the past than
that responsible for the northern plume.

\subsubsection{Supermassive Black Hole Driven Shock} \label{sec:shock}

The third alternative for the origin of the SE edge is a shock driven
by an outburst from the AGN in NGC~6166.  \citet{sf06} highlighted the
complex temperature properties associated with this prominent feature
and argued that they are best explained by an isothermal shock.  There
is clear evidence for AGN outbursts that might have driven a shock,
with the radio lobes and jet excavating cavities in the ICM
(Figs.~\ref{fig:fov}, \ref{fig:central}, and \ref{fig:radio}).  The
relationship between the radio and X-ray features can be seen in the
residual image of Fig.~\ref{fig:central}.  This image shows that the
outer radio lobes and the SE edge have comparable scales.
However, if the shock was initiated by the radio lobes, it is now well
enough detached from them that they are no longer effective in driving
the shock.

The deeper data here generally confirm the observational results of
\citet{sf06}.  The density jump is clear but there is no obvious
temperature jump (Fig.~\ref{fig:front_deproj}).  A density jump of
1.66 would require a Mach number of 1.46 for a conventional shock in
``monatomic'' gas (ratio of specific heats $=5/3$).  Uncertainties in
the deprojected gas properties are dominated by the assumption of
spherical symmetry, since the statistical uncertainties are relatively
small.  The inferred Mach number would imply a temperature jump of
about 1.45, but this is smoothed and reduced when projected onto the
sky.  For example, for the model used to estimate the shock energy
below, the peak of the emission measure weighted temperature is 1.18
times the preshock temperature and it occurs $\simeq 10''$ behind the
shock if the unshocked gas is isothermal.  Binning reduces the
temperature peak further.  No temperature jump is seen, but the
temperature structure of A2199 is complex and the location of the
shock at a radius of $\simeq 100''$ makes detection of the jump
difficult (Fig.~\ref{fig:front_deproj}, left panel).  First, in
contrast to the monotonically declining density distribution, there is
substructure in the gas temperature.  Within $100''$ of the cluster
center, any temperature jump would be superposed on the rising
temperature profile of the cool core in A2199.  This tends to mask a
temperature jump, as found for a shock surrounding M87 in the Virgo
cluster \citep{fjc07}.  By contrast, since the sense of the
temperature jump is the opposite for a cold front, the temperature
gradient would tend to enhance that.  The location of the density jump
in A2199 would place the temperature jump just at the radius of the
transition from a rising to a flat temperature profile
(Fig.~\ref{fig:deproj}).  Second, we cannot know the initial
temperature profile into which the shock was driven, leaving the
temperature baseline unknown --- except by inference from models.
Similar difficulties have been encountered in measuring the
temperature profile of the shock in Hydra A, where temperatures are
affected by the presence of low entropy gas lifted outward by the AGN
outbursts \citep{gnd11}.  Even for the clearest case of a shock in the
very well observed Perseus cluster, no temperature jump has been
detected and it remains unclear why \citep{gfs08, rsf08}.

A2199 shows clear evidence of both repeated AGN outbursts and gas
sloshing resulting from mergers.  The complexities these disturbances
create in its gas distribution may mask the temperature jump that
should accompany the density jump under a shock interpretation.  The
gas distribution has aspherical substructure on scales smaller and
larger than $100''$.  For example, the slope of the temperature
profile steepens beyond a radius of $\simeq 40''$, within which
temperatures are affected by the cool northern plume and low entropy
gas lifted by the radio lobes.  \citet{sf06} argued that the shock is
isothermal, but, while this is consistent with existing data, it is
not required to explain the lack of a clear jump in a system as
complex as A2199.

Like Hydra A, for example \citep{wmn07}, the outburst in Abell~2199
may have created multiple cavities.  The shock in Hydra A closely
envelopes its outer radio lobes \citep{nmw05}, which were only
revealed to be cavities in very deep X-ray images.  Viewed from a
direction more nearly aligned with its radio axis, the outer cavities
of Hydra A would be projected onto the bright cluster center and the
inner cavities, making the outer cavities very difficult to detect.
This would create the appearance of a system like A2199, with shocks
detached from the cavities.  The radius of curvature of the part of
the shock front we would then see in Hydra A is also considerably
larger than its distance from the AGN.  This increases its impact on
the surface brightness, so that, viewed from this direction, the
shocks would also appear stronger, causing us to overestimate the
associated temperature jump.  Although the outer radio lobes of Hydra
A are evident at 300 MHz \citep{lct04}, this is not so at 1.4 GHz,
reminding us that the outer radio lobes may also be ``ghosts'' at
accessible radio frequencies.  By this analogy, if the $100''$ edge in
Abell~2199 is a shock, then we may well expect to find low frequency
radio emission on larger scales than the radio lobes discussed here.

The energy needed to drive a Mach 1.46 shock to a radius of $100''$ in
A2199 can be estimated using a simple spherical shock model.  The
physical model employed here is of a gasdynamic shock initiated by
depositing energy instantaneously at the center of an atmosphere that
was initially hydrostatic and isothermal, with a power-law density
profile \citep[e.g,][]{nmw05}.  Matching the density profile outside
the shock and the density jump at the shock requires a shock energy of
$E_{\rm shock} \simeq 3\times10^{60}$ erg and gives an age of $t_{\rm
  shock} \simeq 25$ Myr, implying a mean power of $\simeq
4\times10^{45}\rm\ erg\ s^{-1}$.  For comparison, we can estimate the
energy needed to create the two large X-ray cavities.  The principle
semi-axes of an ellipse matching the eastern cavity are
$12.1\times8.5$ kpc and those for the western cavity are $14.7\times
9.9$ kpc.  The cavity centers are $\simeq 40''$ from the cluster
center, where the deprojection gives a pressure of $\simeq
1.55\times10^{-10}\rm\ dyne\ cm^{-2}$ (for $kT \simeq 3.2$ keV and
$n_{\rm e}\simeq 0.0157\rm\ cm^{-3}$, Figure~\ref{fig:deproj}).  The
total enthalpy of the two cavities is then $4pV_{\rm tot} \simeq
2.1\times10^{59}$ erg, similar to the value obtained by \citet{rmn06}
using poorer data \citep[our cavity volumes are estimated as the
  geometric mean of the volumes for prolate and oblate ellipsoids, as
  in][]{rmn06}.  While our fitted ellipses do not cover all the
regions with an X-ray deficit, they do generously cover the bulk of
them, so that our enthalpy estimate is unlikely to be very low.  It is
more than an order of magnitude smaller than the shock energy, arguing
against the putative shock being driven by the current radio outburst.
Under adiabatic expansion, the internal energy of a cavity is
proportional to $V^{-(\gamma - 1)}$, where $V$ is the volume and
$\gamma$ is the ratio of specific heats in the cavity.  A radio lobe
dominated by relativistic gas ($\gamma = 4/3$) would have to expand by
a factor of 10 in radius to lose 90\% of its internal energy to its
surroundings, i.e., the shock.  If the shock was launched by the outer
radio lobes, the jet that inflated the lobes must have shut down long
ago, so that, while it was on, the jet power would have been
considerably larger than the estimated mean of $4 \times
10^{45}\rm\ erg\ s^{-1}$ from above.  In the absence of other evident
cavities that could have driven the shock, this may be a weakness of
the shock interpretation.

Note that age estimate of $\simeq 25$ Myr from the explosive shock
model is likely to be low.  Injecting all of the energy at the initial
time in the explosive model maximizes the cavity pressure, hence the
shock speed, at early times.  In more realistic models, the shock is
driven by continuous energy injection, reducing cavity pressures and
the shock speed.  For example, if the shock speed was constant at its
current value of $\simeq 1600\rm\ km\ s^{-1}$, the age of the shock
would be closer to 37 Myr.  The true age could be even longer, since
this does not allow for the temperature decline to the cluster center.
For comparison, the sound crossing time to the centers of the outer
radio lobes at $\simeq 40''$ is $\simeq 25$ Myr.  These estimates
place some constraint on the radio based age estimates for this source
\citep{grt07}.

As discussed above, cold fronts created by sloshing or by remnant
cores are highly asymmetric and generally subtend relatively small
angles at the cluster center.  An additional argument in support of
the shock interpretation for the SE edge is its rather large azimuthal
extent.  Fig.~\ref{fig:central} and Fig.~1 left panel of \citet{sf06}
show the feature extending over at least 180\deg.  Although the radius
and strength of the feature are not constant, this could be explained
by asymmetries in the pre-shock gas distribution \citep[i.e., cluster
``weather,'' e.g.,][]{mhb10}.  The shock speed is modulated by local
gas motions, so that turbulence can produce small scale structure in
an otherwise spherical shock front, smearing out the front when it is
projected onto the sky.  As shown in the Appendix (equation
\ref{eqn:deviation}), turbulence with RMS speed $\vrms$ and coherence
length $\ell$ causes RMS radial displacements in a shock front at
radius $r$ of roughly $\sqrt{r \ell} (\vrms /\vs)$, where $\vs$ is the
shock speed.  Simulations find typical turbulent velocities in the
range 100 -- $300\rm\ km\ s^{-1}$ \citep[e.g.][]{lkn09}, roughly
consistent with observations \citep[e.g.][]{dzw12, sf13}, with
plausible values of the coherence length $\ell \sim 0.1 r$
\citep{rcb05} and scaling with $\vrms$ (equation \ref{eqn:lscale}).
For $\ell = 0.1r$ and $\vrms = 100v_2\rm\ km\ s^{-1}$, this gives
displacements at $r = 100''$ of $\sim 2'' v_2^{3/2}$, so that moderate
levels of turbulence can have a marked smearing effect.  If the SE
edge is a shock, the indistinct appearance of the $100''$ front to the
northwest would then imply that the turbulence is greater there than
on the opposite side of the cluster center.

In conclusion, while the shock interpretation of the SE edge is not
without its problems, we see it as the most viable interpretation of
the observations.

\subsection{The AGN Outburst}

While NGC~6166 hosts an active radio nucleus \citep{grt07}, there are
several indications that the large scale radio source is not as active
now as it has been.  As discussed in \S\ref{sec:shock}, the SE edge is
well detached from the radio lobes and there are no signs of shocks
closer to the lobes.  In particular, the thin shell of dense gas
around parts of the western outer radio lobe is significantly cooler
than the ambient gas at the same distance, $\simeq 60''$, from the
nucleus, with $kT \simeq 2$ keV (Table~\ref{tab:spec}) compared to
$kT\simeq4$ keV for the ambient gas (Figure~\ref{fig:deproj}).  With a
lower temperature and higher density than surrounding gas, this gas
has a lower entropy.  As for similar systems like Hydra A
\citep{ndm02} and Virgo \citep{fnh05}, the low entropy gas would have
been lifted to its present location in the outflow accompanying the
radio outburst.  The lack of shocks close to the radio lobes now shows
that the lobes are expanding subsonically, whereas a considerably more
powerful outburst would have been needed to drive the Mach $\simeq
1.46$ shock that best accounts for the SE edge.

Smaller scale structure associated with the jets confirms this view.
Since the current radio jets appear to be disconnected from the outer
radio lobes, the lobes may be a remnant of an earlier outburst, while
the jets and inner radio lobes are powered by more recent activity.
From \S\ref{sec:shock}, the enthalpy of the outer lobes is $H_{\rm
  lobe} \simeq2.1\times10^{59}$ erg and the sound crossing time to the
the center of these lobes is $t_{\rm lobe} \simeq2.4\times10^7$ yr,
giving a mean jet power to form these lobes of $P_{\rm jet} \simeq
H_{\rm lobe} / t_{\rm lobe} \simeq 2.7\times10^{44}\rm\ erg\ s^{-1}$.
If a comparable power was being injected into the central $10''$ of
the cluster, it would double the thermal energy in $\simeq 1.5
pV/P_{\rm jet} \simeq 1.3$ Myr.  This is several times smaller than
the sound crossing time to $10''$ of $\simeq7.7$ Myr, showing that
such a powerful jet, if confined to the inner $10''$, would drive
strong shocks into the surrounding gas.  The dense rim to the north of
the western jet, at a temperature of $kT \simeq 2$ keV
(Table~\ref{tab:spec}), is no hotter than surrounding gas
(Figure~\ref{fig:deproj}), so it cannot have been subject to a recent
strong shock.  Nor are there any other signs of shocks in this region.
Being so small, estimates like those of \S\ref{sec:shock} for the
enthalpy of the eastern inner lobe are 3 orders of magnitude smaller
than for the outer lobes.

Assuming that the SE edge is a shock, we argued in \S\ref{sec:shock}
that it was produced by a significantly more powerful outburst than
that which produced the current outer radio lobes and cavities.  The
arguments of this section imply that the jet power now is
substantially smaller than the mean power required to create those
features.  If the jet does not penetrate beyond the inner radio
lobes, to explain the lack of strong shocks near the inner radio
lobes requires the power of the jet now to be more than an order of
magnitude smaller than the mean power while inflating the outer radio
lobes.  This implies that the jet power (averaged on time scales of
Myr) has decreased by 2 orders of magnitude or more since the outburst
that produced the $100''$ shock front \citep[cf.][]{wmn07}.

\section{Summary and Conclusions} {\label{sec:summary}}

Deep \chandra{} data for A2199 reveal complex interactions between
radio outbursts from the AGN in NGC~6166 and gas sloshing induced by
past merger activity.  Radio jets from the AGN have excavated cavities
in the X-ray emitting intracluster gas up to 20 -- 30 kpc in diameter
and lifted low entropy gas 30 -- 40 kpc from the cluster center.  The
new data are consistent with previous results that interpret a surface
brightness feature $ 100''$ from the AGN, most evident to the
southeast, as a Mach 1.46 shock front driven by an AGN outburst.
However, we still lack clear evidence of the expected jump in the gas
temperature.  Our estimated range for the age of the shock is 25 -- 37
yr and the energy required to drive it is about $3\times 10^{60}$
erg.

A2199 also shows signatures of gas sloshing induced by merger
activity.  The two clearest signs are excess X-ray surface brightness
$200''$ southwest of the the cluster center and a plume of enhanced
X-ray emission from low entropy, enriched gas that extends $50''$
north of the center.  These resemble spiral enhancements in surface
brightness found in other clusters and in simulations of sloshing
induced by minor mergers, as viewed from roughly the plane of the
merger orbit.  We estimate the time since the core passage that
excited these features as approximately 400 Myr, making them much
older than the shock.  The unusual radio ridge of 3C~338 is associated
with low entropy gas lying to the south of the center of A2199.  This
is consistent with the ridge being the remnant of a former radio jet,
swept southward from the AGN by sloshing gas.  Such an interpretation
requires the magnetic field in the ridge to be somewhat less than our
simple equipartition estimate of $18\ \mu\rm G$.  It also requires the
flow of the sloshing gas to be laminar.

The putative shock is well detached from the outer radio lobes and the
energy required to produce it is an order of magnitude greater than
the enthalpy of the outer lobes.  This suggests that the shock
originated in an earlier outburst than the radio lobes.  There is no
clear connection between the radio jets on scales less than $10''$ and
the outer lobes.  If the jets do not currently penetrate beyond the
inner radio lobes, $7''$ -- $8''$ from the AGN, then the absence of
strong shocks in this region requires the jet power be some two orders
of magnitude smaller now than it was when the southeast shock was
formed.  A2199 is another cluster central radio galaxy showing
evidence of large excursions in jet power on Myr timescales
\citep[e.g.,][]{wmn07, rfg11}.

\acknowledgments

We thank Maxim Markevitch for his assistance and the referee for
helping to improve the paper.  PEJN was supported by NASA grant
NAS8-03060.  The scientific results in this article are based to a
significant degree on observations made with the Chandra X-ray
Observatory.  This research has made use of software provided by the
Chandra X-ray Center in the applications packages CIAO, Chips, and
Sherpa.  The National Radio Astronomy Observatory is a facility of the
National Science Foundation operated under cooperative agreement by
Associated Universities, Inc.  Based on observations made with the
NASA/ESA Hubble Space Telescope, obtained from the Data Archive at the
Space Telescope Science Institute, which is operated by the
Association of Universities for Research in Astronomy, Inc., under
NASA contract NAS 5-26555.  These observations are associated with
program \#9293.

\bibliographystyle{apj}

\bibliography{a2199}

\appendix

\section{Effect of Turbulence on a Shock Front}

The detailed effects of turbulent motion on the propagation of a shock
front are complex, but our purpose is only to estimate the size of
deviations from a spherical shock front.  For that purpose, it is
sufficient to ignore all but the first order effects of turbulent
velocities on the radial velocity of the shock.  If the $x$ component
of the turbulent velocity is $\vtx$, then the velocity of a shock
propagating along the $x$ axis is changed by approximately $\vtx$.  To
first order in $\vtx$, this causes a shock front that would have
reached $r$ to be displaced by
\begin{equation}
\Delta r \simeq \int \vtx \, dt \simeq \int_0^r \vtx \, {dx \over \vs},
\end{equation}
where $\vs$ is the unperturbed shock speed, assumed constant to keep
the argument simple.  We also ignore the time dependence of $\vtx$, on
the assumption that the shock is much faster than $|\vtx|$.  If the
mean turbulent speed is zero, then the variance of $\Delta r$ is
\begin{equation}
\langle (\Delta r)^2 \rangle
= \left \langle \int_0^r \vtx (x') {dx' \over \vs} \int_0^r \vtx (x'')
{dx''\over \vs} \right\rangle
= \int_0^r dx' \int_{-x'}^{r - x'} d\delta {\langle \vtx (x') \vtx (x'
  + \delta) \rangle \over \vs^2},
\end{equation}
where $\langle \rangle$ is the time average and $\delta = x'' - x'$.
Reversing the order of integration, this becomes
\begin{equation}
\langle (\Delta r)^2 \rangle
= \int_{-r}^r d\delta \int_{\max (-\delta, 0)}^{\min (r, r - \delta)} dx'
{\langle \vtx (x') \vtx (x' + \delta) \rangle \over \vs^2}.
\end{equation}
For the purpose of estimating $\langle (\Delta r)^2 \rangle$, we
assume that the turbulence is homogeneous and isotropic (on the scale
of interest), so that the autocorrelation of the $x$ component of the
turbulent velocity, $F(\delta) = \langle \vtx (x') \vtx (x' + \delta)
\rangle$, does not depend on $x'$.  The $x'$ integration then gives $F
(\delta) (r - |\delta|) / \vs^2$ and isotropy implies $F(\delta)$ is
even in $\delta$, so that
\begin{equation}
\langle (\Delta r)^2 \rangle = {2 \over \vs^2} \int_0^r F (\delta) (r
  - \delta) \, d\delta.
\end{equation}
Now $F(0) = \vrms^2 / 3$, where $\vrms$ is the RMS turbulent speed and
$F (\delta) \to 0$ on scales, $|\delta|$, larger than some coherence
length for the turbulence.  Assuming that the coherence length of the
turbulence is small compared to $r$, we get
\begin{equation}
\langle (\Delta r)^2 \rangle \simeq {2\over3} {\vrms^2 \over \vs^2} r \ell,
\qquad {\rm where} \quad 
\ell = \int_0^\infty {F(\delta) \over F(0)} d\delta,
\label{eqn:deviation}
\end{equation}
which is known as an integral scale length for the turbulent velocity,
is a measure of the coherence length of the turbulence.  A similar
estimate is obtained simply by assuming that the front passes through
a number $N = r / \ell$ independent regions with RMS velocities
$\vrms$, so that the RMS deviation in the speed of the shock front as
it propagates to $r$ is $\vrms / \sqrt{N}$.  Displacements, $\Delta
r$, in the shock front will also be correlated on scales up to $\simeq
\ell$.

Because turbulence in the ICM is generally driven by large scale
processes (mergers, AGN outbursts, etc.), $\ell$ will tend to be as
large as possible.  For subsonic turbulence, the upper limit on $\ell$
is generally set by gravity, which opposes the turnover of large
eddies.  In a stably stratified atmosphere, the force per unit mass
opposing the radial displacement from equilibrium of a small fluid
element by $\delta r$ is $\ombv^2 \delta r$, where the \bruntvaisala{}
frequency, $\ombv$, is given by
\begin{equation}
\ombv^2 = {g \over \gamma r} {d\ln\Sigma \over d\ln r} = {\vkepler^2
  \over \gamma r^2} {d\ln\Sigma \over d\ln r}.
\end{equation}
Here, $\gamma$ is the ratio of specific heats for the ICM, $\Sigma$ is
the entropy index (section \ref{sec:center}), $g$ is the acceleration
due to gravity, and $\vkepler$ is the speed of circular orbits.
Equating the radial part of the turbulent kinetic energy per unit
mass, $\vrms^2/6$, to the maximum potential energy per unit mass,
$\ombv^2 \delta r_{\rm max}^2 / 2$, gives the largest radial size of
a turbulent eddy,
\begin{equation}
\delta r_{\rm max} =  \sqrt{\gamma \over 3} {\vrms r \over \vkepler}
\left(d \ln\Sigma \over d \ln r\right)^{-1/2}.
\label{eqn:lscale}
\end{equation}
Typical values of $d \ln\Sigma / d\ln r$ are about unity
\citep[e.g.,][]{cdv09}.  In practice, $\ell$ is somewhat smaller than
$\delta r_{\rm max}$, but it should scale as indicated by equation
(\ref{eqn:lscale}).

\begin{deluxetable}{cccccccc}
\tabletypesize{\footnotesize}
\tablecaption{A Log of {\sl Chandra} observations of Abell 2199}
\tablewidth{0pt}
\tablehead{
\colhead{ObsID} &
\colhead{Date} &
\colhead{Exposure} &
\colhead{RA} &
\colhead{DEC} &
\colhead{Roll} &
\colhead{Detector} &
\colhead{MODE} \\
\colhead{(1)} & 
\colhead{(2)} &
\colhead{(3)} &
\colhead{(4)} &
\colhead{(5)} &
\colhead{(6)} &
\colhead{(7)} &
\colhead{(8)} }
\startdata
\dataset[ADS/Sa.CXO#obs/00498]{498}&
  1999-12-11& 16.2&   247.18789&   39.55308&      11.3&  7& Faint \\
\dataset[ADS/Sa.CXO#obs/00497]{497}&
  2000-05-13&  18.8&   247.13518&   39.55900&      162.3& 7& Faint \\
\dataset[ADS/Sa.CXO#obs/10804]{10804}&
  2009-06-23 & 18.0&   247.13280&   39.53847&      205.6&  0123& VFaint\\
\dataset[ADS/Sa.CXO#obs/10803]{10803}&
  2009-11-17&  28.6&   247.19133&   39.54714&      345.2&  0123& VFaint\\
\dataset[ADS/Sa.CXO#obs/10748]{10748}&
  2009-11-19&  37.7&   247.19134&   39.54714&      345.2&  0123& VFaint\\
\dataset[ADS/Sa.CXO#obs/10805]{10805}&
  2009-11-23&  25.2&   247.19171&   39.54967&       351.1&  0123& VFaint
\enddata
\tablecomments{(1) Observation ID; (2) Date of observation; (3)
  Effective exposure, in ks; (4) \& (5) Celestial coordinates of the
  aimpoint; (6) Roll angle in degrees; (7) CCDs used in this work; (8)
  ACIS data mode.} 
\label{tab:log}
\end{deluxetable}

\begin{deluxetable}{cccccccc}
\tablecaption{Parameters of a beta-model fit to the X-ray
  emission$^a$} 
\tablewidth{0pt} 
\tablehead{ 
\colhead{$R_{\rm in}^b$} &
\colhead{$R_{\rm out}^b$} &
\colhead{$\Delta\alpha^c$} &
\colhead{$\Delta\delta^c$} &
\colhead{$r_0$} &
\colhead{$\beta$} &
\colhead{$\epsilon$} &
\colhead{$\theta$$^d$} } 
\startdata 
$0''$ & $400''$ & $-2''$ & $-4''$ & $27''$ & 0.48 
&0.80 & $37^\circ$ \\ 
$20''$ & $400''$ & $-3''$ & $-5''$ & $37''$ & 0.51 
&0.80 & $37^\circ$ \\ 
$60''$ & $400''$ & $-6''$ & $-8''$ & $46''$ & 0.54 
&0.80 & $38^\circ$ \\ 
$60''$ & $600''$ & $-5''$ & $-8''$ & $53''$ & 0.57 
&0.81 & $39^\circ$ \enddata
\tablecomments{$^a$Fitted with a 2-dimensional beta-model: $f(x,y)
  \propto [1+(r/r_0)^2]^{0.5 - 3 \beta}$, where $r(x,y)^2 = [(x - x_0)
    \cos \theta + (y - y_0) \sin \theta]^2 + [(y - y_0) \cos\theta -
    (x - x_0) \sin \theta]^2 / \epsilon^2 $; $^b$Inner and outer radii
  of the fitted region; $^c$Offset of beta model center from the X-ray
  nucleus; $^d$Position angle, eastward of north.}
\label{tab:beta}
\end{deluxetable}

\begin{deluxetable}{cccccc}
\tabletypesize{\footnotesize}
\tablecaption{
Summary of VLA Observations
}
\tablewidth{0pt}
\tablehead{
 Program code & Observation date & Configuration & Frequency & Beam
 size & Noise \\ 
   &    &   & (GHz)  &  &  ($\mu\rm Jy\ beam^{-1}$)
}
\startdata 
AG269 & 04-JUL-1988 & B & 4.89 GHz &$1.00'' \times 1.00''$ & 3.35\\
AG357 & 02-JAN-1993 & A & 1.49 GHz &$1.30'' \times 1.25''$ & 214
\enddata
\label{tab:radio}
\end{deluxetable}

\begin{deluxetable}{cccccccc}
\tabletypesize{\footnotesize}
\tablecaption{Parameters of selected radio features}
\tablewidth{0pt}
\tablehead{
\colhead{Region} &
\colhead{Size$^a$}  &
\colhead{$S_{\rm 5 GHz}$} &
\colhead{$\alpha$} &
\colhead{$L_\nu$} &
\colhead{$B_{\rm min}$$^b$} &
\colhead{$p_{\rm min}$$^c$} &
\colhead{$t_{\rm syn}$} \\
\colhead{} &
\colhead{(arcsec$^2$)} &
\colhead{(mJy)} &
\colhead{} &
\colhead{($10^{28}{\rm~erg~s^{-1}~Hz^{-1}}$)} &
\colhead{($\mu$G)} &
\colhead{($10^{-11}{\rm~dyn~cm^{-2}}$)} &
\colhead{($10^6$ yr)}
}
\startdata
Western Inner Lobe & $\pi\times1^2$ & 1.3 & -1.0 & 2.5 & 21.0 & 1.4 & 5.5\\
Eastern Inner Lobe & $\pi\times1^2$ & 1.6 & -1.1 & 3.1 & 22.2 & 1.6 & 5.0\\
Ridge        & 15$\times$3 & 16.3 & -1.3 & 31.5 & 18.0 & 1.0 & 7.6
\enddata
\tablecomments{$^a$The line-of-sight depth is assumed to equal to the
  width of each feature; $^b$Minimum energy condition, assuming the
  power law electron distribution corresponding to $\alpha$ between
  cutoffs corresponding to 10 MHz and 100 GHz, with only radiating
  particles and magnetic field; $^c$Relativistic electron plus
  magnetic pressure for the minimum energy condition.}
\label{tab:radio2}
\end{deluxetable}

\begin{deluxetable}{ccccccc}
\tabletypesize{\footnotesize}
\tablecaption{Measurements of the H$\alpha +$[N II] emission}
\tablewidth{0pt}
\tablehead{
\colhead{Region$^a$} &
\colhead{Centroid RA} &
\colhead{Centroid Dec} &
\colhead{Area} &
\colhead{Luminosity$^b$} &
\colhead{Gas density$^c$} &
\colhead{Mass} \\
\colhead{} &
\colhead{(J2000)} &
\colhead{(J2000)} &
\colhead{($\rm arcsec^2$)} &
\colhead{($10^{39}\rm\ erg\ s^{-1}$)}&
\colhead{($f^{-0.5}\rm\ cm^{-3}$)} &
\colhead{($f^{0.5} 10^6\rm\ M_\odot$)} } 
\startdata
Nucleus &\fmtra{16 28 38.26} & \fmtdec{39 33 04.4}& $\pi \times 1^2$ & 45.4 &
  0.068 & 1.3 \\ 
Knot &\fmtra{16 28 38.46} & \fmtdec{39 33 04.3} & $\pi \times 1^2$ & 6.9 &
  0.026 & 0.5\\ 
Filament$^d$ &\fmtra{16 28 37.94} & \fmtdec{39 33 03.2} & $4 \times0.4$ & 
  1.1 & 0.033 & 0.6 
\enddata 
\tablecomments{$^a$Regions described in text; $^b$Assuming a
  conversion factor of $1.24 \times 10^{-18}
  \rm\ erg\ cm^{-2}\ s^{-1}$ \AA$^{-1}/(\rm ct\ s^{-1})$ and an FWHM
  of 130 \AA; $^c$For case B recombination at $10^4$ K, with volume
  filling factor $f$. To estimate the volume, the line-of-sight depth
  is assumed to be equal to the width of the region; $^d$Long axis
  $50^\circ$ east of north.}
\label{tab:Halpha}
\end{deluxetable}

\begin{deluxetable}{cccccccc}
\tabletypesize{\footnotesize}
\tablecaption{Spectral fits for various regions}
\tablewidth{0pt}
\tablehead{
\colhead{Region$^a$} &
\colhead{$N_{\rm H}^b$} &
\colhead{$kT$} &
\colhead{$Z$} &
\colhead{$\chi^2/\rm d.o.f.$} &
\colhead{Volume} &
\colhead{$n^c$} &
\colhead{Flux$^d$} \\
\colhead{ } & 
\colhead{($10^{20}\rm\ cm^{-2}$)} &
\colhead{(keV)} &
\colhead{($Z_\odot$)} &
\colhead{} &
\colhead{(${\rm cm}^3$)} &
\colhead{($\rm cm^{-3}$)} &
\colhead{($10^{-14}\rm\ erg\ cm^{-2}\ s^{-1}$)}
}
\startdata
Shell & 0.9 & 1.93$^{+0.55}_{-0.30}$ & 0.6 (fixed) & 101/101 &1.2$\times10^{67}$& 0.042&3.3 \\ 
Rim & 5.0 (fixed) &1.98$^{+0.63}_{-0.35}$&0.68 (fixed)  & 37/46 &3.0$\times10^{65}$&0.20&1.8\\ 
Knot & 0.9 & 1.00$^{+0.12}_{-0.13}$ & 0.13$^{+0.10}_{-0.06}$ & 35/37 &1.5$\times10^{65}$&0.42&2.1
\enddata
\tablecomments{$^a$See Figs.~\ref{fig:central} and \ref{fig:core}.
  CIAO format region for shell: $\mathop{\rm pie}
  (16\mathord:28\mathord:34.50, +39\mathord:33\mathord:04.6, 10",
  22", -30, 30)$, shell background: $\mathop{\rm pie}
  (16\mathord:28\mathord:34.50, +39\mathord:33\mathord:04.6, 22",
  34", -30, 30)$, rim: $\mathop{\rm rotbox}
  (16\mathord:28\mathord:37.92, +39\mathord:33\mathord:07.4, 6", 3",
  0)$, rim background: $\mathop{\rm rotbox}
  (16\mathord:28\mathord:37.92, +39\mathord:33\mathord:10.3, 6", 3",
  0)$, knot: $\mathop{\rm rotbox} (16\mathord:28\mathord:38.52,
  +39\mathord:33\mathord:04.3, 3", 3", 0)$, knot background:
  $\mathop{\rm rotbox} (16\mathord:28\mathord:38.52,
  +39\mathord:33\mathord:07.2, 3", 3", 0)$; $^b$Fixed at the Galactic
  foreground value for the shell and knot, and at the best fit for the
  local background for the rim ($kT$ and $n$ are very insensitive to
  these values of $N_{\rm H}$); $^c$From the model normalization,
  assuming that the line-of-sight depth equals the width of each
  region; $^d$0.5 -- 8 keV unabsorbed flux. }
\label{tab:spec}
\end{deluxetable}

\end{document}